\documentclass[twocolumn,showpacs,preprintnumbers,amsmath,amssymb]{revtex4}
\usepackage{epsfig}

\usepackage{graphicx}
\usepackage{dcolumn}
\usepackage{bm}
\usepackage{axodraw}

  \usepackage{pstricks}
  \usepackage{amssymb,amsmath}

\newpsobject{ast}{psdot}{dotstyle=asterisk,dotsize=5pt}
\def\ea{{\it et al.}}

\newcommand{\ex}[1]{{\begin{picture}(13,0)(0,0)\put(0,0){\rm
#1}\put(0,0){\line(2,1){15}}\end{picture}}~~~~~}
\newcommand{\exx}[1]{{\begin{picture}(13,0)(0,0)\put(0,0){\rm
#1}\put(0,0){\line(2,1){15}}\end{picture}}~}

\newcommand{\beq}{\begin{equation}}
\newcommand{\eeq}{\end{equation}}
\newcommand{\beqa}{\begin{eqnarray}}
\newcommand{\eeqa}{\end{eqnarray}}
\def\lsim{\ \rlap{\raise 3pt \hbox{$<$}}{\lower 3pt \hbox{$\sim$}}\ }
\def\gsim{\ \rlap{\raise 3pt \hbox{$>$}}{\lower 3pt \hbox{$\sim$}}\ }

\def\fv{\mathbf{5}}
\def\bfv{\mathbf{\bar{5}}}
\def\fu{\mathbf{5}_{\phi_u}}
\def\bfd{\mathbf{\bar{5}}^{\phi_d}}
\def\t{\mathbf{ 10}}
\def\bt{\mathbf{\bar{10}}}
\def\fif{\mathbf{ 15}}
\def\bfif{\mathbf{\bar{15}}}
\def\tfv{\mathbf{ 35}}
\def\btfv{\mathbf{\bar{35}}}
\def\forty{\mathbf{ 40}}
\def\bforty{\mathbf{\bar{40}}}
\def\for{\mathbf{ 45}}
\def\bfor{\mathbf{\bar{45}}}
\def\fifty{\mathbf{ 50}}
\def\bfifty{\mathbf{\bar{50}}}
\def\se{\mathbf{ 70}}
\def\bse{\mathbf{\bar{70}}}

\def\tf{\mathbf{24}}

\begin{document} 

\preprint{}

\title{ 
 Fermion mass hierarchy and non-hierarchical
 mass ratios in~$\mathbf{SU(5)\times U(1)_F}$
}

 \author{Luis F. Duque,$^{a,b}$ 
         Diego A. Gutierrez,$^a$
         Enrico Nardi$^{a,c}$ 
         and Jorge Nore\~na$^{d}$ }

\affiliation{\vspace{2mm}
\small${}^a$Instituto~de~F\'{i}sica,~Universidad~de~Antioquia, A.A.{\it{1226}} Medell\'\i n,~Colombia
\vspace{1mm} 
\\${}^b$ 
ITM,~Calle~73~No.76A-354~Medell\'{\i}n,~Colombia
\vspace{1mm} \\
            ${}^c$INFN-Laboratori Nazionali di Frascati,  
                   C.P. 13, I-00044 Frascati, Italy 
\vspace{1mm}\\
             ${}^d$SISSA/ISAS, I-34013 Trieste, Italy
}


\begin{abstract} 
\noindent
We consider a $SU(5)\times U(1)_F$ GUT-flavor model in which the number of
effects that determine the charged fermions Yukawa matrices is much larger
than the number of observables, resulting in a hierarchical fermion spectrum
with no particular regularities.  The GUT-flavor symmetry is broken by flavons
in the adjoint of $SU(5)$, realizing a variant of the Froggatt-Nielsen
mechanism that gives rise to a large number of effective operators.  By
assuming a common mass for the heavy fields and universality of the
fundamental Yukawa couplings, we reduce the number of free parameters to one.
The observed fermion mass spectrum is reproduced thanks to selection rules
that discriminate among various contributions.  Bottom-tau Yukawa unification
is preserved at leading order, but there is no unification for the first two
families.  Interestingly, $U(1)_F$ charges alone do not determine the
hierarchy, and can only give upper bounds on the parametric suppression of the
Yukawa operators.
\end{abstract} 

\maketitle



\section{Introduction}
 \label{sec:introduction}
\vspace{-3mm}
The Standard Model (SM) provides and accurate description of particle physics
phenomena. Particles interactions are derived from local symmetries and are
explained at a fundamental level by the gauge principle. Myriads of
experimental tests have confirmed the correctness of this picture.  However,
the SM cannot explain the values of the particle masses and mixing angles, and
to get insight into this issue a new theory is required.  

The mass spectrum of the charged fermions has a strong hierarchical structure
that ranges over five orders of magnitude. Apart from this obvious feature,
only a few regularities are observed, the most certain of which is that the
bottom and tau masses converge towards a similar value when extrapolated to
some large energy scale~$\sim 10^{16}\,$GeV.  In the context of the MSSM, a
 recent analysis finds~\cite{Ross:2007az}
\beq
\frac{m_b}{m_\tau}= 1.00^{+0.04}_{-0.4}.  
\label{eq:btau} 
\eeq
Supersymmetric unification of the three gauge couplings occurs at the same
energy scale, and this hints to a grand unified theory (GUT) that can explain
elegantly this result. It is thus likely that $m_b$-$m_\tau$ unification is not
a numerical accident (although it  could be only an approximate
result~\footnote{The number quoted correspond to $\tan\beta = 1.3$. For
  larger $\tan\beta$ and neglecting 
  threshold corrections, a smaller value $m_b/m_\tau\approx 0.73\pm0.03$ is
  found~\cite{Ross:2007az}.}).  For the down-quark and leptons of the first
two generations unification does not work, but a different GUT relation exists:
$3m_s/m_\mu=m_d/3m_e=1$. This relation was suggested long ago by Georgi and
Jarlskog (GJ)~\cite{Georgi:1979df} that also showed how they could be obtained
in the context of $SU(5)$ by means of a $\for$ Higgs representation.  The GJ
relations are less certain: the analysis in~\cite{Ross:2007az} quotes (for
$\tan\beta =1.3$~\footnote{For larger $\tan\beta$ ref.~\cite{Ross:2007az}
  quotes $3m_s/m_\mu \approx 0.69 \pm 0.08$ while $m_d/3m_e$ is practically
  unaffected.}):
\beq
\frac{3m_s}{m_\mu}= 0.70^{+0.8}_{-0.05}, \qquad \frac{m_d}{3m_e}=0.82 \pm 0.07.  
\label{eq:GJ} 
\eeq
It is then likely that a more complicated mechanism is responsible for the
values of these mass ratios. As regards the oldest mass matrix relation
$V_{us}\approx \sqrt{m_d/m_s}$ that was proposed forty years ago by Gatto,
Sartori and Tonin (GST)~\cite{Gatto:1969dv}, it is still in good agreement
with the experimental values~\cite{Ross:2007az}. A few other empirical
relations were proposed in~\cite{Ferrandis:2004ti}.

The absence of enough well established regularities in the fermion mass
pattern leaves open the way to many different explanations of the origin of
fermion masses, and is probably the main reason why, in spite of all the
theoretical efforts, no compelling theory has yet emerged.  Most theoretical
efforts concentrated in reducing the number of fundamental parameters as much
as possible, by imposing symmetries and/or by assuming special textures for
the Yukawa matrices, like symmetric forms, or a certain number of zero
elements (for reviews of different ideas see e.g.
refs.~\cite{Raby:1995uv,Hall:1993nr,Berezhiani:1995tr,Chen:2003zv}).  Clearly,
a number of parameters smaller than the number of observables would yield some
testable predictions, that can rule out some possibilities and favor others.
Moreover, there is also the hope that a reduced set of parameters could reveal
some regular pattern that could provide some hint of the correct theory.

However, it is also possible that the opposite situation is true.  Namely,
that the number of different effects that contribute to determine the values
of the fermion masses is much larger than the number of observables.  Then,
even if the fundamental contributions are determined by some simple rule or
symmetry principle, it is likely that in the fermion spectrum no regularities
would appear.  If this is the case, and if the scale of the related new
physics is inaccessibly large, then identifying the correct solution to the
fermion mass problem could be an impossible task.

In this paper we discuss a framework that realizes this possibility.  We
assume the GUT-flavor symmetry $SU(5)\times U(1)_F$, that is broken
down to the SM by $SU(5)$ adjoint Higgs representations charged under
$U(1)_F$, that hence play also the role of flavon fields.  The
Froggatt-Nielsen (FN) mechanism~\cite{Froggatt:1978nt} is incorporated with
the variant that since the flavons are not singlets under $SU(5)$, the heavy
vectorlike fields responsible for generating the effective mass operators for
the quarks and leptons can belong to several different representations.  To
simplify things, and to highlight the special features of this framework, we
assume that all the heavy states have the same mass, and that at the
fundamental level the Yukawa couplings are universal. This yields a scheme
with just one relevant parameter, that is the ratio between the vacuum
expectation value (vev) of the flavons and the heavy fermions mass. This
parameter is responsible for the fermion mass hierarchy, while the details of
the spectrum are determined by several non-hierarchical (and computable) group
theoretical coefficients, that depend on the way the heavy FN states are
assigned to $SU(5)$ representations.  As we will see, the number of
contributions to the fermion mass operators, each one weighted by a different
$SU(5)$ coefficient, overwhelms the number of observable, completely hiding
the underlying $SU(5)$ symmetry.

\vspace{-2mm} 
\section{The general framework}
\vspace{-2mm} 
We work in the framework of a supersymmetric GUT-flavor model
based on the gauge group $SU(5)\times U(1)_F$ (supersymmetry is needed for
$SU(5)$ to be a phenomenologically viable GUT).  Different realizations of
models based on $SU(5)\times U(1)_F$ have been proposed expecially for what
concerns the implications for the possible patterns of neutrino masses and
mixing angles (see \cite{altarelli} for a review and list of references).  In the
following we list the main ingredients and assumptions that underlie our 
framework.\\  [-2pt]

{\it $SU(5)$ Grand Unified Symmetry}. \ 
In $SU(5)$ GUTs, the $SU(2)$ lepton doublets $L=(\nu,e)^T$ and the down-quark
singlets $d^c$ are assigned to the fundamental conjugate representation
$\mathbf{\bar 5}$, while the quark doublets $Q=(u,\,d)^T$, the up-type quark
singlets $u^c$ and the lepton singlets $e^c$ fill up the two-index antisymmetric
$\mathbf{10}$:
\begin{equation}
  \label{eq:fiveten}
  \bfv = 
\begin{pmatrix}
d^c\\[-1pt]d^c\\[-1pt]d^c\\[-1pt] e\\[-1pt] -\nu
\end{pmatrix}\!,
\quad\!\!
  \t = \frac{1}{\sqrt{2}} 
\begin{pmatrix}
\!0\!   &\! u^c\!&\!-u^c\!&\! u \! &d\!\\[-1pt]
\!-u^c\!&\!  0\! &\!u^c\! & \!u \! &\!d\!\\[-1pt]
\!u^c\! &\!-u^c\!&\! 0\!  &\! u \! &d\!\\[-1pt] 
\!-u\!  &\! -u\! &\! -u\! &\! 0  \!&e^c\!\\[-1pt] 
\!-d\!  &\! -d\! &\! -d\! &\!-e^c\!&\!0\!
\end{pmatrix}\!.
\end{equation}
The Higgs field $\phi_d$ responsible for the down-quarks and lepton masses
belong to another $\mathbf{\bar 5}^{\phi_d}$ with $\langle \mathbf{\bar
  5}^{\phi_d}\rangle \sim {\rm diag} (0,0,0,0,-v_d)$, while $\phi_u$
responsible for the masses of the up-quarks is assigned to a fundamental
$\mathbf{5}_{\phi_u}$ with $\langle \mathbf{5}_{\phi_u}\rangle \sim {\rm diag}
(0,0,0,0,v_u)$. As usual we define $\tan\beta\equiv v_u/v_d$.  The Yukawa
superpotential is
\begin{equation}
\label{eq:WY}
 W_Y \!=\! 
\sqrt{2}\,Y_{IJ}^D\> \mathbf{\bar 5}_{Ia}
\mathbf{10}_{J}^{ab} \, \mathbf{\bar 5}^{\phi_d}_b + 
\frac{1}{4}Y_{IJ}^U\> \epsilon_{abcde} \mathbf{10}_I^{ab} \mathbf{10}_J^{cd} 
\,\mathbf{5}_{\phi_u}^e,  
\end{equation}
where $Y^D$ ($Y^U$) is the Yukawa couplings matrix for the down-quarks and
leptons (up-quarks), $I\,,J=1,2,3$ are generation indices,
$a,b,\dots=1,\dots,5$ are $SU(5)$ indices, and $\epsilon_{abcde}$ is the
$SU(5)$ totally antisymmetric tensor.

One problem of $SU(5)$ GUTs is how to guarantee that the two electroweak Higgs
doublets $H_u$ and $\bar H_d$ contained respectively in $\fu$ and $\bfd$
remain light, while the color triplet components acquire a mass large enough
to suppress proton decay below the experimental limits.  The technical
solution adopted in minimal $SU(5)$ is to invoke a fine tuned cancellation
between a trilinear term coupling $\fu$ and $\bfd$ with the adjoint $\Sigma$ 
(with vev $\langle \Sigma\rangle =V\, {\rm diag}(2,2,2,-3,-3)$) and an
invariant mass term:
\begin{equation}
  \label{eq:splitting}
W_\phi \sim {\bfd}_a (\Sigma^a_b + 3 M_\phi \delta^a_b){\fu}^{\!\!\!\!\!\!b}\
\> .   
\end{equation}
By choosing $M_\phi = V$ with an accuracy of one part in $10^{14}$ the
contribution to the SM Higgs doublets is of the order of $100\,$GeV,
while the color triplets acquire a GUT scale mass. \\

{\it $U(1)_F$ flavor symmetry and FN mechanism}.\ Two qualitative features are
apparent in the (GUT scale) charged fermion mass spectrum: $i)$ the structure
is strongly hierarchical; $ii)$ there is no obvious inter-family multiplet
structure.  The first feature hints to a spontaneously broken flavor symmetry
in which the hierarchical structure is determined by powers of a small
order parameter, while the second feature suggests that the
flavor symmetry is likely to contain an Abelian factor \cite{Leurer:1993gy}.

The approach proposed long ago by Froggatt and Nielsen \cite{Froggatt:1978nt}
realizes these two conditions.  The basic ingredient is an Abelian flavor
symmetry that forbids at the renormalizable level most of the fermion Yukawa
couplings. The symmetry is spontaneously broken by the vev of a SM singlet
flavon field $\langle S\rangle$. After the symmetry is broken a set of
effective operators arises, that couple the SM fermions to the electroweak
Higgs boson(s), and that are induced by heavy vectorlike fields with mass $M >
\langle S\rangle$.  The hierarchy of fermion masses results from the
dimensional hierarchy among the various higher order operators that are
suppressed by powers of the ratio $\langle S \rangle/M<1$.  In turn, the
suppression powers are determined by the Abelian charges assigned to the
fermion fields.  This mechanism has been thoroughly studied in different
contexts like the supersymmetric SM \cite{Leurer:1993gy}, in frameworks where
the horizontal symmetry is promoted to a gauge symmetry that can be anomalous
\cite{Ibanez:1994ig,Binetruy:1994ru,GaugeU1} or
non-anomalous~\cite{Mira:1999fx}, and with discrete Abelian
symmetries~\cite{Plentinger:2008up}.  When incorporated in the SM (or in the
MSSM~\cite{Leurer:1993gy}) the FN mechanism allows to account qualitatively
for the hierarchy in the fermion mass pattern and can also yield a couple of
order-of-magnitude predictions.  However, when applied to the simplest GUT
models, like those based on $SU(5)$, the FN mechanism is less successful.  On
the one hand $U(1)_F$ breaking by $SU(5)$ singlet flavons does not account for
the mass ratios $m_s/m_\mu,\, m_d/m_e\neq 1$. On the other hand, the fact that
the five fermion multiplets for generation of the SM are reduced to just two
$SU(5)$ representations eq.~(\ref{eq:fiveten}) implies much less freedom in
choosing the $U(1)_F$ charges, and generally only the gross features of the
fermion mass spectrum can be accounted for.

As we will discuss, these drawbacks can be overcome if the flavon fields are
assigned to $SU(5)$ adjoint representations.  Thus we assume that the same
scalar multiplets that break $SU(5)$ down to the SM gauge group carry flavor
charges and break also $U(1)_F$, playing effectively the role of the singlet
flavon $S$ in usual FN models.  At each new order in the (small) symmetry
breaking parameter a cascade of new effective operators appears.  These
operators are weighted by non-trivial $SU(5)$ group theoretical factors, and
contribute differently to the down-quark and lepton mass matrices. This allows
to explain the lifting of the mass degeneracy between leptons and quarks
belonging to the same multiplet while, rather surprisingly, under certain
conditions approximate $b$-$\tau$ unification can be preserved (see also 
ref.~\cite{daristi}). \\

{\it Universal Yukawa interactions and heavy fermion masses.}  We implement
the variant of the FN mechanism described above within a quite constrained
framework, considering the possibility that the fermion mass structure could
result from an underlying model in which at a high scale (larger than the GUT
scale) the fundamental Yukawa couplings obey to some unification principle, 
analogous to the unification found for the gauge couplings.  This assumption
will be treated just as a constraining condition that, thanks to the large
reduction in the number of free parameters, allows to highlight some general
features of the model. In particular we will not speculate on the origin of
this universality~\footnote{While there are theoretical frameworks in which
  Yukawa couplings obey to some principle of universality, like e.g.
  superstring-inspired models~\cite{kobayashi}, SUSY-GUT models for
  gauge-Yukawa unification~\cite{mondragon}, or models for gauge-Higgs
  unification in higher dimensions \cite{burdman}, the fundamental Yukawa
  couplings of our model involve rather large vectorlike representations that
  cannot be accommodated easily in these scenarios.}.  We will also assume a
common mass ($M > \Lambda_{GUT}$) for all the heavy vectorlike representations
(unlike the previous assumption, this condition can be implemented rather
easily by assuming that the vectorlike masses are dominated by the common vev
of a singlet scalar field).  With these two assumptions there remains only one
free parameter that is relevant to the problem, that is the dimensionless
ratio between the symmetry breaking vev and the heavy mass $M$.\\

In summary, the scheme we are proposing embeds the FN explanation of the
hierarchy of the fermion mass spectrum, but introduces additional group
theoretical structures. They have a twofold effect: firstly they can change
the naive hierarchy that one would infer from the $U(1)_F$ charge assignments
(and could even produce texture zeros); secondly, they result in a large set
of non hierarchical coefficients that depend on the field content of the
model, and that can simulate rather well the absence of regular
patterns in the effective Yukawa matrices. The important point is
that these coefficients are computable, and we will illustrate their effects
by evaluating the lepton and down-type quark Yukawa matrices up to third order
in the small symmetry breaking parameter.  

\section{$U(1)_F$ charge assignments}
\vspace{-2mm} In this section we discuss a set of conditions that can help us
to identify the possible charge assignments for the SM fields.
\subsection{Fermion  mass hierarchy} 
\vspace{-2mm} A reasonable description of the charged fermion mass hierarchy
can be given in terms of the following mass ratios (we assume from the start
moderate values of $\tan\beta$):
\begin{eqnarray}
  \label{eq:ratiosdl}
m_{d,\,e}\,:\>m_{s,\,\mu}\,:\>m_{b,\,\tau}&\approx& \epsilon^3\,:\> 
\epsilon^2\,:\>\epsilon, \\ 
  \label{eq:ratiosu}
  m_u\,:\>m_c\,:\>m_t &\approx& \epsilon^4\,:\> \epsilon^2\,:\>1,\qquad 
\end{eqnarray}
with $\epsilon \approx 1/20 - 1/30$. 
These relations imply that the order of magnitude of the determinants
of the Yukawa matrices is 
\begin{equation}
  \label{eq:det}
  \det Y^U \sim \det Y^D\sim  \epsilon^6.
\end{equation}
We now assume that $\epsilon$ is related to the vev of flavon fields that,
without loss of generality, carry a unit charge under the $U(1)_F$ symmetry.
More precisely, we assume that the flavor symmetry is broken by scalar fields
$\Sigma_{\pm1}$ in the $\mathbf{24}$-dimensional adjoint representation of
$SU(5)$, where the subscripts $\pm1$ refer to the $U(1)_F$ charge values, and
set the normalization for all the other charges.  The vevs $\langle
\Sigma_{+1}\rangle=\langle \Sigma_{-1}\rangle = V_a$ (as required by $D$
flatness) with $V_a=V\, {\rm diag}(2,2,2,-3,-3)/\sqrt{60} $ are also
responsible for breaking the GUT symmetry down to the electroweak-color gauge
group.  The order parameter for the flavor symmetry is then $\epsilon=V/M$
where $M$ is the common mass of the heavy FN vectorlike fields.  This symmetry
breaking scheme has two important consequences:
\begin{enumerate} \itemsep=3pt
\item Power suppressions in $\epsilon$  appear 
with coefficients  related  to the different entries in $V_a$, 
that distinguish the leptons from the quarks.
\item The FN fields are not restricted to the $\fv,\bfv$ or $\t,\bt$
  multiplets as is the case when the $U(1)_F$ breaking is triggered by singlet
  flavons.
\end{enumerate}

After the symmetry is broken, the effective Yukawa couplings generated
for the charged fermions are suppressed {\it at least} as
\begin{equation}
\label{eq:suppression}   
Y^D_{IJ} \sim \epsilon^{|\bfv_I+\t_J+\bfd|}, \qquad 
Y^U_{IJ} \sim \epsilon^{|\t_I+\t_J+\fu|}, 
\end{equation}
where we denote the $F$-charges with the same symbol than the multiplets, e.g.
$F(\bfv_I)=\bfv_I$, $F(\t_I)=\t_I$, etc.  Since we have two flavon multiplets
$\Sigma_{\pm 1}$ with opposite charges, the horizontal symmetry allows for
operators with charges of both signs, and hence the exponents in
eq.~(\ref{eq:suppression}) are absolute values of sum of charges.
For models based on spontaneously broken family symmetries it is natural to
assume that the mass hierarchies (\ref{eq:ratiosdl}) and (\ref{eq:ratiosu}) are
determined by the diagonal terms in the Yukawa matrices, that is that the
off-diagonal terms are smaller than the respective combinations of on-diagonal
matrix elements.  This gives for the multiplet charges the following
conditions:
\begin{eqnarray}
  \label{eq:firstcond} 
\nonumber
\hspace{0cm}  
|\bfv_3+\t_3+\bfd| =1, &\hspace{.3cm}&  |\t_3+\t_3+\fu|=0,\qquad \\ 
|\bfv_2+\t_2+\bfd|=2,  &             &  |\t_2+\t_2+\fu|=2, \\ \nonumber
|\bfv_1+\t_1+\bfd|=3,  &             &  |\t_1+\t_1+\fu|=4. 
\end{eqnarray}
Since the top-quark mass is of the order of the electroweak breaking scale,
$Y^U_{33}$ must be the fundamental coupling of a renormalizable operator.
Namely, the top Yukawa coupling must respect the flavor symmetry.  All the
other mass operators are not $U(1)_F$ invariant, and the corresponding Yukawa
couplings are effective parameters suppressed by powers of $\epsilon$.

\subsection{Distinguishing fermion multiplets at the fundamental level.}
\vspace{-2mm}
When a pair of multiplets $\mathbf{r_1}$ and $\mathbf{r_2}$ are assigned to
the same representation of the fundamental gauge group (in our case
$SU(5)\times U(1)_F$), the gauge-interaction Lagrangian has a global $U(2)$
symmetry corresponding to rotations of $\mathbf{r_{1,2}}$. If also the Yukawa
Lagrangian respects this global symmetry, one combination of the two
multiplets decouples and remains massless. Since no massless fermions are
observed, if this kind of symmetries exist, they must be broken.  In the
absence of any other fundamental `label' that would distinguish $\mathbf{r_1}$
from $\mathbf{r_2}$, the only possibility is to introduce an explicit
breaking.  In Abelian models of flavor this breaking is usually provided {\it
  ad-hoc} by assuming that different ${\cal O}(1)$ coefficients multiply the
Yukawa terms for $\mathbf{r_1}$ and $\mathbf{r_2}$. However, distinguishing
$\mathbf{r_1}$ from $\mathbf{r_2}$ can be also considered as part of the
flavor problem, and this part is left unexplained by the ad-hoc procedure.  Of
course one can assume that the different ${\cal O}(1)$ coefficient have an
explanation at a more fundamental level, but this still implies that an
additional (unspecified) structure able to distinguish $\mathbf{r_1}$ from
$\mathbf{r_2}$ must exist.  Therefore, one is forced either to give up the
possibility of a full explanation of the flavor problem, or to assume that the
model is incomplete at least in some parts.

Because of the assumption of universality of the fundamental Yukawa couplings,
we cannot appeal to ${\cal O}(1)$ coefficients of unspecified origin, and
accordingly we will require that each fermion multiplet is assigned to a
different $SU(5)\times U(1)_F$ representation.  Note that this conditions
excludes the simple (and often used) charge assignments in which the
hierarchical patterns (\ref{eq:ratiosdl}), (\ref{eq:ratiosu}) are reproduced
by assigning the same charge to all the $\bfv_I$, and the hierarchy is
determined by augmenting in each generation the charge of $\t_I$ by one unit
(see \cite{Antebi:2005hr} for a theoretical framework that could explain such
a pattern of $U(1)_F$ charges).

\subsection{Doublet-triplet splitting}
\vspace{-2mm} 
Implementing the technical solution to the doublet-triplet splitting problem
eq.~(\ref{eq:splitting}) within our model implies two conditions.  Firstly, to
allow for the invariant mass term $M_\phi$ the charges of the multiplets
containing the Higgs fields must be equal in magnitude and opposite in sign:
\begin{equation}
  \label{eq:higgs}
  \bfd+\fu=0.
\end{equation}
Secondly, we need to introduce an adjoint Higgs representation $\Sigma_0$
neutral under $U(1)_F$ to implement the cancellation between the two
contributions to the  Higgs doublets mass.  Note that replacing $\Sigma$
in eq.~(\ref{eq:splitting}) by an effective term $\Sigma_+\Sigma_-/M$ would
suppress the masses of the color triplets by one power of $\epsilon$, and this
could imply an unacceptably fast proton decay.  While our model does not shed
any new light on the origin of the doublet-triplet splitting, the two
conditions above must be imposed to ensure its (technical) consistency.

\subsection{Charge assignments}
\vspace{-2mm} 
The requirement that all the SM fermion multiplets are assigned
to inequivalent $SU(5)\times U(1)_F$ representations together with the
conditions eqs.~(\ref{eq:firstcond}) and (\ref{eq:higgs}) result in eight
possible charge assignments that are listed in table~\ref{table:1} where, for
simplicity, we have arbitrarily chosen vanishing charges for the Higgs fields
$\bfd=\fu=0$.  (Reverting the sign of all charges gives trivially other eight
possibilities.)

The effective Yukawa superpotential eq.~(\ref{eq:WY}) is invariant with
respect to the following charge redefinitions:
\begin{eqnarray}\nonumber 
\bfv_I &\to& \bfv_I\  +\   a; \\  \nonumber 
\t_I & \to& \t_I+b; \\  \label{eq:redefinitions}  
\bfd & \to& \bfd-(a+b);\\ \nonumber 
\fu &\to& \fu-2\, b, \nonumber 
\end{eqnarray}
with $a$ and $b$ two arbitrary real numbers.  Under this redefinitions the
charge of the Higgs bilinear term shifts as
\begin{equation}
  \label{eq:bilinear}
  F(\bfd\fu) \to  F(\bfd\fu) -(a+3b), 
\end{equation}
and thus for $b=-a/3$ the redefinitions eq.~(\ref{eq:redefinitions}) leave
invariant also the condition eq.~(\ref{eq:higgs}).  Therefore, more precisely
each row in table~\ref{table:1} identifies a one-parameter family of charges
satisfying eqs.~(\ref{eq:firstcond}) and~(\ref{eq:higgs}).  The assignments in
the last six rows (3)-(8) in the table must be discarded at once since they
yield mass eigenvalues for the leptons and down-quarks of order $\sim v_d$.
This is because in all these cases for some entries of the down and lepton
Yukawa matrix $Y^D$ the sum of the charges vanishes, and thus these entries
are not suppressed by powers of $\epsilon$.  This does not happen for the
first two assignments, that could lead to viable models.
\begin{table}[h] 
  \centering \tabcolsep4.5pt
   \begin{tabular}{|c||r|r|r||r|r|c|}\hline
\phantom{$Big|$}        & $\bfv_1$& $\bfv_2$ &$\bfv_3$
& $\t_1$ & $\t_2$ & $\t_3$
\\ \hline\hline
(1) & $-5 $& $-3 $ &$+1 $& $+2 $& $+1 $ & $ 0 $ \\ \hline
(2) & $+5 $& $-3 $ &$+1 $& $-2 $& $+1 $ & $ 0 $\\ \hline
(3) &  $+1 $& $-3 $ &$-1 $& $ +2 $& $ +1 $ & $ 0 $ \\ \hline
(4) & $+5 $& $-3 $ &$-1 $& $-2 $& $ +1 $ & $ 0 $ \\ \hline
(5) & $-5 $& $-3 $ &$-1 $& $ +2 $& $ +1 $ & $ 0 $  \\ \hline
(6) & $-1 $& $-3 $ &$+1 $& $-2 $& $ +1 $ & $ 0 $ \\ \hline
(7) & $-5 $& $+1 $ &$-1 $& $ +2 $& $ +1 $ & $ 0 $  \\ \hline
(8) & $+5 $& $+1 $ &$-1 $& $ -2 $& $+1 $ & $0 $ \\ \hline
   \end{tabular}
  \caption{
The eight $U(1)_F$ charge assignments that satisfy the hierarchy 
conditions in eq.~(\ref{eq:firstcond}) together with eq.~(\ref{eq:higgs}), and 
that  label in a distinguishable way 
all the fermion multiplets. In all the cases $\bfd=\fu=0$ has been  
chosen for simplicity.  
 }
  \label{table:1}
\end{table}
\subsection{Gauge anomalies}
\vspace{-2mm} 
We assume that the $SU(5)\times U(1)_F$ symmetry is gauged, and thus it
must be free of gauge anomalies.  Anomaly cancellation yields three
conditions corresponding to the vanishing of the gravitation-$U(1)_F$ anomaly,
of the pure $U(1)_F^3$ anomaly, and of the mixed $SU(5)^2\times U(1)$ anomaly.
The first two anomalies can be canceled by adding $SU(5)$ singlet states with
suitable $U(1)_F$ charges (e.g. two singlet `neutrinos' \cite{Chen:2008tc}).
The mixed anomaly can be canceled by invoking the Green-Schwarz mechanism
\cite{Green:1984sg}, in which case the $U(1)_F$ symmetry is called {\it
  anomalous}, or by satisfying the condition
\begin{equation}
{\cal A} \equiv 
{\cal I}_{\mathbf 5}
\left(\bfd + \fu + \sum_{I=1}^3  \bfv_I  \right) +  
{\cal I}_{\mathbf{10}}  \sum_{I=1}^3 \t_I = 0,
\label{eq:anomalies}  
\end{equation}
in which case the symmetry is {\it non-anomalous}.  In
eq.~(\ref{eq:anomalies}) ${\cal I}_{\mathbf{5}}$ and ${\cal I}_{\mathbf{10}}$
denote the index of the respective representations: ${\rm
  Tr}[\mathbf{r}^a\mathbf{r}^b] \equiv {\cal I}_{\mathbf{r}}\,\delta^{ab}$
($\mathbf{r}=\fv,\,\t$), while $\bfd,\,\fu,\,\bfv_I,\, \t_I$ denote the
$U(1)_F$ charges.  Condition (\ref{eq:higgs}) implies that the Higgs
representations $\bfd$ and $\fu$ do not contribute to ${\cal A}$. Then, under
the charge redefinitions (\ref{eq:redefinitions}) restricted to $b=-a/3$ to
preserve (\ref{eq:higgs}), the anomaly coefficient shifts as
\begin{equation}
\label{eq:shift} 
{\cal A} \to {\cal A}' = {\cal A} +3a\,\left({\cal
  I}_{\mathbf{5}}-
\frac{{\cal I}_{\mathbf{10}}}{3}\right).
\end{equation} 
It is a numerical coincidence that the ratio of the indexes for the $\t$ and
$\bfv$ $SU(5)$ representations is ${\cal I}_{\mathbf{10}}/{\cal
  I}_{\mathbf{5}}=3$~\cite{Slansky:1981yr} implying that the remaining one
parameter freedom in redefining the charges is ineffective for canceling the
mixed anomalies. More precisely, no charge redefinition is possible that
cancels the mixed anomaly and simultaneously preserves eq.~(\ref{eq:higgs}).
By inspecting the first two rows in table~\ref{table:1} we can then conclude
that the assignments in (1) correspond to a family of anomalous models, and 
those in (2) to a family of anomaly free models.\\

The one parameter freedom in charge redefinition can still be useful to forbid
all the trilinear operators $\bfv_I\bfv_J \t_K$ that violate baryon and lepton
number, as well as the lepton number violating bilinear terms $\bfv_I\fu$.  To
achieve this we shift the charges of the first set in table~\ref{table:1} with
$a=-3b=+1$, and the charges of the second set with $a=-3b=-1$. This yields the
charge assignments given in table~\ref{table:2}.  With these assignments,
R-parity arises as an accidental symmetry enforced by $SU(5)\times U(1)_F$
gauge invariance.
 \begin{table}[h] 
  \centering \tabcolsep3.5pt
   \begin{tabular}{|c||r|r|r||r|r|r||c|}\hline
 \phantom{$Big|$} &$\bfv_1$& $\bfv_2$ &$\bfv_3$& $\t_1$ & $\t_2$ & $\t_3$
         &$\footnotesize\fu\!=\!-\bfd$ \\ \hline\hline
(1)&     $-4 $& $-2 $ &$2 $& $5/3 $ & $2/3 $ & $-1/3 $ &$2/3$ \\ \hline
(2)& $-4 $& $4 $ &$0 $& $5/3 $ & $-4/3 $ & $-1/3 $ &$2/3$ \\ \hline
   \end{tabular}
   \caption{
     The charge assignments (1) and (2) of 
     table~\ref{table:1} redefined to forbid  the R-parity 
     violating couplings $\bfv_I\bfv_J \t_K$ and 
     $\bfv_I\fu$ by means of the shifts in eq.~(\ref{eq:redefinitions}) with 
     $a=-3b=+1$ for case (1) and $a=-3b=-1$ for case (2).
}
  \label{table:2}
\end{table}

\section{The Model}
\vspace{-2mm} For the two sets of $U(1)_F$ charges in table~\ref{table:2},
charge counting suggests the following (naive) hierarchical patterns for the
Yukawa matrices.  For the anomalous case (1) we have
\begin{equation} 
\label{eq:Y1}
Y^D\sim
 \left(
  \begin{array}{ccc}
   \epsilon^{3} & \epsilon^{4} & \epsilon^{5}\\
   \epsilon^{} & \epsilon^{2} & \epsilon^{3}\\
   \epsilon^{3} & \epsilon^{2} & \epsilon^{}
  \end{array}
 \right),
  \qquad
Y^U\sim 
 \left(
  \begin{array}{ccc}
   \epsilon^{4} & \epsilon^{3} & \epsilon^{2}\\
   \epsilon^{3} & \epsilon^{2} & \epsilon^{}\\
   \epsilon^{2} & \epsilon^{} &  1
  \end{array}
 \right), 
\end{equation}
where in $Y^D$ the rows correspond to $(\bfv_1,\bfv_2,\bfv_3)$ and the
columns to $(\t_1,\t_2,\t_3)$. The order of magnitude of the
determinants of $Y^D$ and $Y^U$ is
\begin{equation}
\label{eq:det1}
\det Y^D \sim 
\det Y^U \sim \epsilon^6.
\end{equation}
For the non-anomalous case (2) we have 
\begin{equation} 
\label{eq:Y2}
Y^D \sim 
 \left(
  \begin{array}{ccc}
   \epsilon^{3} & \epsilon^{6} & \epsilon^{5}\\
   \epsilon^{5} & \epsilon^{2} & \epsilon^{3}\\
   \epsilon^{1} & \epsilon^{2} & \epsilon^{1}
  \end{array}
 \right)
  \qquad
Y^U\sim 
 \left(
  \begin{array}{ccc}
   \epsilon^{4} & \epsilon^{} & \epsilon^{2}\\
   \epsilon^{} & \epsilon^{2} & \epsilon^{}\\
   \epsilon^{2} & \epsilon^{} &  1
  \end{array}
 \right), 
\end{equation}
that gives 
\begin{equation}
\label{eq:det2}
\det Y^D\sim \epsilon^6, \qquad  
\det Y^U \sim \epsilon^2.
\end{equation}
In case (1) the order of magnitude of the determinants eq.~(\ref{eq:det1})
agrees with the phenomenological result eq.~(\ref{eq:det}).  In contrast, for
the second set of charges, $\det Y^U$ in eq.~(\ref{eq:det2}) is four powers of
$\epsilon$ too large.  This does not necessarily imply that case (2) is not
viable. As we will show, in the present framework charge counting only gives
upper bounds on the parametrically suppressed couplings, and it is still
possible that $Y^U_{12,21}$ in eq.~(\ref{eq:Y2}) could be promoted to ${\cal
  O}(\epsilon^3)$ thus recovering $\det Y^U\sim \epsilon^6$.  Nevertheless, in
the rest of this paper we will concentrate on the anomalous case (1) that
looks phenomenologically more promising, and that will allow us to illustrate
all the interesting features of our scheme while dealing just with the
down-quarks and leptons Yukawa matrix $Y^D$.

Since the form of $Y^U$ in eq.~(\ref{eq:Y1}) is approximately diagonal, the
up-quark mass hierarchy is be accounted for by the diagonal entries, while
off-diagonal entries contribute to the quark mixing matrix.  Given that
experimental uncertainties for the GUT scale up-quarks mass ratios are rather
large, ($m_u/m_c \sim 0.002-0.004$, $m_c/m_t \sim 0.001 - 0.003$, see
\cite{Ross:2007az,Fusaoka:1998vc}) we assume that these ratios can be
accommodated within the model, and we concentrate on the structure of $Y^D$.
Reconciling the matrix $Y^D$ in eq.~(\ref{eq:Y1}) with experimental
observation appears quite challenging, firstly because of an apparent
`anomaly' in the (2,1) entry that spoils approximate diagonality (this entry
is crucial since it controls the value of the Cabibbo angle) and secondly
because satisfying the down-quarks and leptons mass relations within $SU(5)$
GUTs is not a trivial task.  In the following we analyze the contributions of
different effective operators to $Y^D$, showing that a phenomenologically
acceptable structure,  able to reproduce (approximately) the correct mass
ratios and to give reasonable quark-mixings can be recovered.

\subsection{Effective operators}
\vspace{-2mm}
We assume that a large number of vectorlike FN fields exist in various $SU(5)$
representations.  Since the mass $M$ of these fields is assumed to be larger
than $\Lambda_{GUT}$, at this scale the contributions to the down-quarks and
leptons mass operator $\sim \mathbf{\bar 5}_{Ia} \, \mathbf{10}_{J}^{ab} \>
\mathbf{\bar 5}^{\phi_d}_b$ can be evaluated by means of insertions of
pointlike propagators. We denote the contraction of two vectorlike fields in
the representations $\mathbf{R}$, $\mathbf{\bar R}$ as
\begin{equation}
  \label{eq:pointlike}
  [\mathbf{R}^{abc\dots}_{de\dots}\mathbf{\bar
  R}_{lmn\dots}^{pq\dots}]=
\frac{-i}{M} {\cal S}^{abc\dots pq\dots}_{de\dots lmn\dots},
\end{equation}
where all the indices are $SU(5)$ indices, and ${\cal S}$ is the appropriate
group index structure.  The structures ${\cal S}$ for several $SU(5)$
representations are given in appendix~\ref{sec:group}.  We further use
$\langle \Sigma_\pm \rangle= (V/\sqrt{60})\times {\rm diag}(2,2,2,-3,-3)$
where the factor $1/\sqrt{60}$ gives the usual normalization of the $SU(5)$
generators ${\rm Tr}(\mathbf{R}^a \mathbf{\bar R}^b)=\frac{1}{2}\delta^{ab}$.
The number of effective operators rapidly increases with increasing
powers of $\epsilon$: we find 4 possible operators at ${\cal O}(\epsilon)$, 17
at ${\cal O}(\epsilon^2)$, and more than 70 at ${\cal O}(\epsilon^3)$ (see
appendix~\ref{sec:tables}). 
To establish a path in this forest of operators, we will stick to the
following rule: {\it At each order in $\epsilon$, all the possible operators
  that can arise are allowed to contribute, unless there is a compelling
  reason to forbid specific contributions} \footnote{At ${\cal O}(\epsilon^3)$
  we omit a few operators induced by representation with dimension $> 100$. We
  also omit one operator induced by the $\fifty$ since deriving its group
  index structure is rather awkward.}.  To forbid one operator, we simply
assume that the representations that gives rise to it does not exist.

The assumptions of a unique heavy mass parameter $M$ and of universality of
the fundamental Yukawa coupling (whose value can be absorbed in the vev $V$,
except for an overall factor that cancels in the mass ratios) implies a high
level of predictivity.  Of course there are still several sources of
theoretical uncertainties, and before getting into numbers let us briefly
discuss the most relevant ones.

\begin{itemize}\itemsep=0pt 
\item[{\it i) }] In general, different FN representations
  $\mathbf{R},\,\mathbf{R^\prime},\dots$ can mix through terms like
  $\mathbf{R}\, \Sigma\, \mathbf{R^\prime}$ thus splitting the heavy mass
  eigenstates.  Formally this is an effect of relative order $\epsilon$.
  
\item[{\it ii) }] Each entry in the Yukawa matrices receives contributions
  from higher order operators involving insertions of $\Sigma_0$. Assuming
  $\langle \Sigma_0 \rangle \sim \langle \Sigma_\pm\rangle$ also these
  corrections are of relative order $\epsilon$.
  
\item[{\it iii) }] Universality for the fundamental Yukawa coupling holds at a
  scale $>\Lambda_{GUT}$.  However, for different $SU(5)$ representations the
  renormalization group (RG) evolution down to $\Lambda_{GUT}$ differ, and
  since rather large representations are often involved, RG effects can
  effectively split the GUT scale  couplings.
\end{itemize}
With increasing powers of $\epsilon$, larger group theoretical coefficients
and larger FN representations are involved, and all the corrections listed
above grow, and can become larger than naive estimates.  Therefore, we
should expect that for the lighter fermions the results will be less precise.
However, note that once the field content of a model is specified, all the
corrections listed above are in principle computable.

\subsection{Bottom-tau Yukawa unification}
\label{sec:btau}
\begin{figure}[t]
 \begin{center}
\includegraphics[width=8.5cm,height=2.5cm,angle=0]{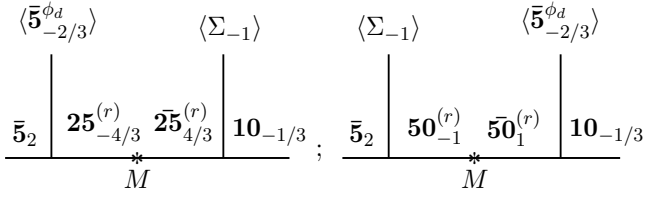}
 \caption[]{
Diagrammatic representation of the ${\cal O}(\epsilon)$ contributions to
$Y^D_{33}$. The  subscripts give the $U(1)_F$ charges.}
 \label{fig1}
\end{center}
\end{figure}

\vspace{-2mm} The $\tau$ and $b$ masses arise at ${\cal O}(\epsilon)$ and to a
good approximation are determined by the values of $Y^D_{33}$.  We will denote
as $Y^\ell_{IJ}$ and $Y^d_{IJ}$ the values of $Y^D_{IJ}$ respectively for the
leptons and quarks, since in general they differ.  The tensor products
relevant to identify the FN representations involved in the $b$-$\tau$ mass
operator are
\begin{eqnarray} 
\label{reducible1}
\bfv \otimes \bfd  &=&\mathbf{\bar{25}}^{(r)}=\bt\oplus\bfif\,, \\ 
\label{reducible2}
\t \otimes \bfd  &=& \mathbf{50}^{(r)}=\fv\oplus  \for.  
\end{eqnarray}
Some results are obtained more easily in terms of reducible representations
like $\mathbf{\bar{25}}^{(r)}$ and $\mathbf{50}^{(r)}$. We use a superscript
$^{(r)}$ to denote the reducible character of a representation and avoid
confusion with irreducible representations of equal dimensionality.  In
eqs.~(\ref{reducible1}) and~(\ref{reducible2}) the second equalities gives the
irreducible fragments.  The representations conjugate to the tensor products
in eqs.~(\ref{reducible1}) and~(\ref{reducible2}) are found respectively in
the tensor products of $\t$ and $\bfv$ with the $\tf$ adjoint
\begin{eqnarray}
\t \otimes \tf &=&\mathbf{25}^{(r)}\oplus\mathbf{40}\oplus\mathbf{175}\,, \\
\bfv \otimes \tf &=& \mathbf{\bar{50}}^{(r)} \oplus \se\,. 
\end{eqnarray}
Hence, at ${\cal O}(\epsilon)$ the contributions to the $b$-$\tau$ mass
operator involve $ \mathbf{25}^{(r)}$, 
$\mathbf{50}^{(r)}$ 
and their conjugate representations, as is diagrammatically depicted in
fig.~\ref{fig1}.
%
%
In terms of irreducible representations, the following operators arise:
\begin{eqnarray}
\label{eq:Obtau10}
O(\epsilon;\t_{-4/3})&=& \bfv_a\bfd_b \left[{\mathbf{10}}^{\,ab}\
\mathbf{\bar {10}}_{lm}\right]\Sigma^m_n\t^{nl}, \qquad 
\\
\label{eq:Obtau15}
O(\epsilon;\fif_{-4/3})&=& 
\bfv_a\bfd_b \left[{\mathbf{15}}^{\,ab}\
\mathbf{\bar {15}}_{lm}\right]\Sigma^m_n\t^{nl},  
\\
\label{eq:Obtau5}
O(\epsilon;\ \fv_{-1})&=&  \bfv_a\Sigma^a_b\,\left[{\fv}^{b}\ 
{\bfv}_{l}\right]\bfd_m\t^{lm},  
\\
\label{eq:Obtau45}
O(\epsilon;\for_{-1})&=& 
  \bfv_a\Sigma^c_b\,\left[ {\for}^{\,ba}_c\ 
{\bfor}^{n}_{ml}\right] \bfd_n\t^{lm}. 
\end{eqnarray}
We denote as $O(\epsilon^n;\ {\mathbf{R'}}_{F_1}, \mathbf{R''}_{F_2},\dots
\mathbf{R^{\it n}}_{F_n} )$ an $SU(5)$ invariant operator of order
$\epsilon^n$ (i.e. with $n$-insertions of $\Sigma$) induced by a
set of $n$ pairs of FN fields in the representations $\mathbf{R}$, with charge
$F$. The string of $\mathbf{R}$'s is ordered from left to right assuming
always that the $\bfv$ and $\t$ containing the SM fields are respectively the
first and last element of the string. Note that together with the values of
the $F$-charges, this determines univocally the order of  insertion of
$\bfd_{-2/3}$ and $\Sigma_{\pm1,0}$.

Using the relevant group structures ${\cal S}$ and the vertices ${\cal V}$
given in appendix~\ref{sec:group}, restricting $\bfd$ to the $SU(2)$ Higgs
doublet (indices $i=4,5$) and projecting $\Sigma$ on the vacuum $\propto {\rm
  diag}(2,2,2,-3,-3)$ we obtain the contributions to $Y^\ell_{33}$ and
$Y^d_{33}$ given in table~\ref{table:Yd33}.
\begin{table}[h] 
  \centering \tabcolsep4.5pt
   \begin{tabular}{|c|r|r|}
\hline 
\phantom{$^\big|$}&$\frac{Y^\ell_{33}}{\epsilon/\sqrt{60}}$&$\frac{Y^d_{33}}{\epsilon/\sqrt{60}}$\\ [2pt] \hline
\phantom{$^\big|$}$O(\epsilon;\t_{-4/3}  )$   & $6 $   & $1 $   \\ 
\phantom{$\big|$}$O(\epsilon;\fif_{-4/3})$  & $ 0 $   & $5 $   \\ [2pt] \hline
\phantom{$^\big|$}$O(\epsilon;\ \fv_{-1})$     & $-3 $   & $2 $  \\ 
\phantom{$\big|$}$O(\epsilon;\for_{-1})$   & $15$   & $10$  \\ [2pt] \hline  
\hline
\phantom{$\Big|$}$\sum_\mathbf{R}O(\epsilon;\mathbf{R})$     
                     & $18$   & $18$ \\ [2pt]
\hline
   \end{tabular}
   \caption{
Coefficients of the  four  operators  that can  contribute  to
$Y^\ell_{33}$ and $Y^d_{33}$   at order $\epsilon$, in units of 
$\epsilon/\sqrt{60}$. The equality of the two sums  given in the 
last row corresponds to  $b$-$\tau$ Yukawa unification.}
  \label{table:Yd33}
\end{table}
It is a non-trivial result that summing up the four contributions for the
leptons and for the quarks one obtains the same result. This allows us to
conclude that {\it the most general effective operator allowed at order
$\epsilon$ by the $SU(5)\times U(1)_F$ symmetry yields $b$-$\tau$ Yukawa
unification}.

Note that unification is already preserved by the individual contributions of
the two reducible representations: $O(\epsilon;\mathbf{25}^{(r)})=
O(\epsilon;\t ) + O(\epsilon;\fif) \to 6 $ and $ O(\epsilon;\mathbf{50}^{(r)})
=O(\epsilon;\ \fv ) + O(\epsilon;\for) \to 12$. This result can be easily
verified by using the ${\cal S}$ structures for the reducible representations
given in appendix~\ref{sec:group} eqs.~(\ref{twentyfive}) and~(\ref{fifty}):
%
\begin{eqnarray}
\bfv_a\bfd_b \left[{\mathbf{25}^{(r)}}^{\,ab}\ 
\mathbf{\bar {25}}^{(r)}_{lm}\right]\Sigma^m_n\t^{nl}   \to  
\frac{6\epsilon}{\sqrt{60}}\, \bfv_a  \t^{ai}  \bfd_i, \quad\ \  &&
\\     
\bfv_a\Sigma^c_b\, \left[{\mathbf{50}^{(r)}}^{ba}_c\
{\mathbf{\bar {50}}}^{(r)n}_{ml}\right]\bfd_n\t^{lm}    \to   
\frac{12 \epsilon}{\sqrt{60}}\, \bfv_a \t^{ai}\bfd_i.  \quad\   &&
\end{eqnarray}
where $i=4,5$ corresponds to the $SU(2)$ components of $\bfd$.  We see that in
both cases the coefficients are independent of the particular value of the
index $a=1,2,3$ ($b$-quark) or $a=4$ ($\tau$-lepton)\footnote{The fact that
  the sum of $O(\epsilon;\t )$ and $O(\epsilon;\fif)$ preserves $b$-$\tau$
  unification was already noted in~ref.~\cite{daristi}.}.  Therefore, in terms of
contributions of reducible representations, unification is preserved even if
the assumption of universality of the Yukawa couplings is relaxed.  Note
however, that this is only a leading order result, since in general the ${\cal
  O}(\epsilon^2)$ corrections to $Y^\ell_{33}$ and $Y^d_{33}$ will be
different. Therefore, the models predicts that $b$-$\tau$ Yukawa unification
is only an {\it approximate leading order result}.

From table~\ref{table:Yd33} we see that for the $\tau$-lepton the effective
operator involving the $\fif$ has a vanishing coefficient.  This happens
because the $\fif$ is symmetric in $SU(5)$ indices while the $\t$ is
antisymmetric. The $SU(2)$ singlet leptons correspond to $\t^{45,54}$.
Specifying these indices forces $\langle \Sigma \rangle$ to the lowest
$2\times 2$ corner, that is proportional to the identity $I_{2\times 2}$, and
thus because of the symmetric-antisymmetric contraction the operator vanishes.
It is not difficult to prove that the same happens also for higher order
diagrams containing the $\fif$.  Similar implications of $SU(5)$ index
properties occur also for other representations. In particular, the effective
operators involving the $\forty$ also vanish for the leptons.
  
\subsection{The $Y^D_{21}$ entry and the Cabibbo angle}
\vspace{-2mm} 
$U(1)_F$ charge counting suggests that the (2,1) entry in $Y^D$
eq.~(\ref{eq:Y1}) can arise already at ${\cal O}(\epsilon)$.  This would spoil
the approximate diagonality of $Y^D$ and induce a too large mixing between the
down-quarks of the first and second generations. In short, such a parametric
suppression is too mild to be phenomenologically acceptable. In our scheme
this entry can be generated by operators analogous to the ones in
eqs.~(\ref{eq:Obtau10})-(\ref{eq:Obtau45}) but with FN fields with different
charges: $O(\epsilon;\t_{8/3})$, $O(\epsilon;\fif_{8/3})$,
$O(\epsilon;\fv_{+1})$ and $O(\epsilon;\for_{+1})$. Therefore, if such
representations do not exist, at order $\epsilon$ the $Y^D_{21}$ entry
vanishes.  At the next order it could be generated by means of one
insertion of the neutral adjoint $\Sigma_0$.  The same restrictions that
forbid the ${\cal O}(\epsilon)$ terms also imply that only two operators are
allowed: $O(\epsilon^2;\se_{+1},\fif_{+5/3})$ and
$O(\epsilon^2;\for_{+2},\bforty_{+8/3})$.  $\se_{+1}$ must exist in order to
generate $Y^D_{22}$ at ${\cal O}(\epsilon^2)$ (see below), therefore to
forbid the first operator we assume that $\fif_{+5/3}$ is absent.  As regard
the second operator, the absence of $\bforty_{+8/3}$ would forbid it, but
would also imply a strong suppression of the $Y^D_{11}$ entry (see below).
This could even be a welcome feature, since a highly suppressed (or texture
zero) $Y^d_{11}$ is one of the conditions required to obtain the
GST~\cite{Gatto:1969dv} relation.  However, to stick to our guiding principle
of allowing for the maximum possible number of contributions (and to avoid
regularities) we assume instead that $\for_{+2}$ is absent. With these two
assumptions, $Y^D_{21}$ vanishes also at ${\cal O}(\epsilon^2)$ and
approximate diagonality for $Y^D$ is recovered.  In summary, by assuming that
no FN fields exist with quantum numbers corresponding to $\fif_{8/3},\,
\t_{8/3},\, \fv_{+1},\, \for_{+1},\, \fif_{+5/3} $ and $\for_{+2}$, the
$Y^D_{21}$ entry is promoted to ${\cal O}(\epsilon^3)$. The possible operators
that can contribute at this order are listed in table~\ref{tab:Yd21}.
%
\begin{table}[h!!] 
\begin{tabular}{|l|r|r|}
\hline
  \multicolumn{3}{|l|}{
  \phantom{$\big|$}\ex{$\t_{+8/3}$,}~~\ex{$\fif_{+8/3}$,}~~\ex{\
  $\fv_{+1}$,}\ex{$\for_{+1}$;}\ex{$\fif_{+5/3}$,}~~\ex{$\for_{+2}$}  }  \\
\hline 
 \phantom{$\Big|^l$} &$\frac{Y_{21}^\ell}{(\epsilon/\sqrt{60})^3}$ & $\frac{Y_{21}^d}{(\epsilon/\sqrt{60})^3}$ \\
\hline
 \phantom{$\Big|$}$\left[\Sigma_{-1}\bfd_{-2/3}\Sigma_{+1}\Sigma_{+1}\right]^{\phantom{l}
  }$ & 
& \\ 
 \phantom{$\big|$}$O(\epsilon^3;\fv_{+3},\t_{+11/3},\bforty_{+8/3})$&$ 0 $&$ 100 $ \\
 \phantom{$\big|$}$O(\epsilon^3;\for_{+3},\t_{+11/3},\bforty_{+8/3})$&$ 0 $&$  500$ \\
 \phantom{$\big|$}$O^\uparrow(\epsilon^3;\for_{+3},\bforty_{+11/3},\bforty_{+8/3})$&$0$&  $800$  \\
 \phantom{$\big|$}$O^\downarrow(\epsilon^3;\for_{+3},\bforty_{+11/3},\bforty_{+8/3})$&$0$&$-1400$ \\
 \hline
 \phantom{$\Big|$}$\left[\Sigma_{+1}\Sigma_{+1}\bfd_{-2/3}\Sigma_{-1}\right]^{\phantom{l}}$ & 
& \\ 
\phantom{$\big|$}$O(\epsilon^3;\se_{+1},\mbox{\exx{$\
    \fv_0$}},\mbox{{$\t_{+2/3}$}})$ & \exx{$1350$}~$\>$&\qquad \exx{$200$}~\\
\phantom{$\big|$}$O(\epsilon^3;\se_{+1},\mbox{\exx{$\ \fv_0$}},\fif_{+2/3})$ & $0$&\exx{$1000$}~~\\
\phantom{$\big|$}$O(\epsilon^3;\se_{+1},\for_0,\mbox{{$\t_{+2/3}$}})$ & $-1350$&{$200$}~\\
\phantom{$\big|$}$O(\epsilon^3;\se_{+1},\for_0,\bforty_{+2/3})$&$0$&$4000$ \\
\phantom{$\big|$}$O^\uparrow(\epsilon^3;\se_{+1},\se_0,\fif_{+2/3})$&$0$&$-400 $ \\
\phantom{$\big|$}$O^\downarrow(\epsilon^3;\se_{+1},\se_0,\fif_{+2/3})$&$0$&$800$ \\
 \hline
 \phantom{$\Big|$}$\left[\Sigma_{-1}\Sigma_{+1}\bfd_{-2/3}\Sigma_{+1}\right]^{\phantom{l}}$ & 
& \\ 
\phantom{$\big|$}$O(\epsilon^3;\for_{+3},\mbox{\exx{$\bfifty_{+2}$}\ },\bforty_{+8/3})$  &$0$& $-$\\ [1pt]
\hline
\hline
\phantom{$\Big|$}$\sum_\mathbf{R}O(\epsilon^3;\mathbf{R})$     
              & $-1350$   & $4600$ \\ [2pt]
\hline
\end{tabular}
   \caption{Contributions to $Y^D_{21}$  at ${\cal O}(\epsilon^3)$ 
in units of $(\epsilon/\sqrt{60})^3$. The upper row lists the 
representations in the absence of which $Y^D_{21}$ is promoted to ${\cal
  O}(\epsilon^3)$.  The three possible orders of  insertion of $\bfd$ and $\Sigma$'s
are specified in square brackets. 
Constraints from $Y^D_{22}$ imply crossing out $\bfv_0$, while 
the  $\bfifty$  is crossed out because it is not included in the analysis.  
The sums do not include the crossed out coefficients.
}  
  \label{tab:Yd21}
\end{table}
Note that no operators involving a pair of $\Sigma_0$ is allowed.  Note also
that when the $\forty\Sigma\bforty$ and $\bse\Sigma\se$ vertices are involved,
two different contributions are present (respectively, the third and fourth,
ninth and tenth entries in table~\ref{tab:Yd21}).  This is because for these
vertices two inequivalent contractions of the $SU(5)$ indices are possible
(see eqns.~(\ref{v40})-(\ref{v70}) in appendix~\ref{sec:group}).  This is
always the case when a representation is contained twice in its tensor product
with the adjoint $\mathbf{R}\otimes \Sigma = \mathbf{R}\oplus \mathbf{R}\oplus
\dots$ ($ \mathbf{R}=\bforty,\,\for,\,\se\dots$).  We distinguish the two
contributions by means of up- and down-arrow labels
$O^\uparrow,\,O^\downarrow$.


%
\vspace{-.1cm}
\subsection{The $Y^D_{22}$ entry and the strange and muon masses}
\vspace{-2mm} 
After the FN representations are restricted according with the previous
discussion, at ${\cal O}(\epsilon^2)$ only two contributions to $Y^D_{22}$
remain possible.  They are given in table~\ref{tab:Yld22eps2}.
%
\begin{table}[h!!] 
\begin{tabular}{|l|r|r|}
\hline
 \phantom{$\Big|^l$} &$\frac{Y_{22}^\ell}{(\epsilon/\sqrt{60})^2}$ & $\frac{Y_{22}^d}{(\epsilon/\sqrt{60})^2}$ \\
\hline
 \phantom{$\Big|$}$\left[\Sigma_{+1}\Sigma_{+1}\bfd_{-2/3}\right]^{\phantom{l}}$ & 
& \\ 
\phantom{$\big|$}$O(\epsilon^2;\se_{+1},\mbox{\exx{$\ \fv_0$}})$ &
$-$\exx{$\ 225\> $}\, & $-$\exx{$\ 200\ $}\  \\
\phantom{$\big|$}$O(\epsilon^2;\se_{+1},\for_0)$ & $225$&$-200$ \\
\hline
\end{tabular}
   \caption{Contributions to $Y^D_{22}$  at ${\cal O}(\epsilon^2)$ 
in units of $(\epsilon/\sqrt{60})^2$. According to the discussion in the text, 
the operator involving the $\fv_0$ has been  crossed out.
}  
  \label{tab:Yld22eps2}
\end{table}
We see that for the leptons a cancellation occurs. 
(This cancellation can also be traced back to the symmetry/antisymmetry in the
two upper indices respectively of the $\se$ and of the reducible
$\fifty^{(r)}=\fv\oplus\for$.)  If both contributions are allowed, $m_\mu$
would be formally suppressed to ${\cal O}(\epsilon^3)$. Therefore we assume
that $\fv_0$ is absent, and we keep only the contribution of the $\for_0$.
Having estimated $Y^\ell_{33}$ and $Y^\ell_{22}$, we can now fit $\epsilon$
to the value of the $\mu$ and $\tau$ mass ratio $Y^\ell_{33}/Y^\ell_{22}\sim
m_\mu/m_\tau\sim 0.06$ obtaining
\begin{equation}
\label{eq:epsilon}
\epsilon \sim \frac{18\sqrt{60}}{225}\,\frac{m_\mu}{m_\tau} \simeq 0.037.
\end{equation} 
The ratio between the top and bottom Yukawa couplings can be estimated using
this value of $\epsilon$:
\begin{equation}
\frac{Y^U_{33}}{Y^d_{33}}\sim \left(\frac{18\epsilon}{\sqrt{60}}\right)^{-1}  
\simeq 12. 
\end{equation} 
Note that this result points towards moderate values of $\tan\beta$
($\tan\beta < 10$~\cite{Ross:2007az}).  From the sum for the contributions in
table~\ref{tab:Yd21} and from table~\ref{tab:Yld22eps2} we can also see that
the down-quark sector gives a sizeable contribution to Cabibbo mixing
$Y^d_{21}/Y^d_{22} \sim 0.11$.
%
\begin{table}[h!!] 
\begin{tabular}{|l|r|r|}
\hline
 \phantom{$\Big|^l$} &$\frac{Y_{22}^\ell}{(\epsilon/\sqrt{60})^3}$ & $\frac{Y_{22}^d}{(\epsilon/\sqrt{60})^3}$ \\
\hline
\phantom{$\Big|$}$\left[\Sigma_{+1}\Sigma_{+1}\bfd_{-2/3}\Sigma_{0}\right]^{\phantom{l}}$ & 
& \\ 
\phantom{$\big|$}$O^\uparrow(\epsilon^3;\se_{+1},\se_{0},\fif_{+2/3})$ & $0$&$-400$ \\
\phantom{$\big|$}$O^\downarrow(\epsilon^3;\se_{+1},\se_{0},\fif_{+2/3})$&$0$&$800$ \\
\phantom{$\big|$}$O(\epsilon^3;\se_{+1},\for_{0},\t_{+2/3})$&$ -1350 $&$ 200$ \\
\phantom{$\big|$}$O(\epsilon^3;\se_{+1},\for_{0},\bforty_{+2/3})$&$ 0 $&$ 4000 $ \\
 \hline
 \phantom{$\Big|$}$\left[\Sigma_{0}\Sigma_{+1}\Sigma_{+1}\bfd_{-2/3}\right]^{\phantom{l}}$ & 
& \\ 
\phantom{$\big|$}$O(\epsilon^3;\fv_{+2},\se_{+1},\for_{0})$ & $675$&$400$ \\
\phantom{$\big|$}$O^\uparrow(\epsilon^3;\se_{+2},\se_{+1},\for_{0})$ & $4725$&$800$ \\
\phantom{$\big|$}$O^\downarrow(\epsilon^3;\se_{+2},\se_{+1},\for_{0})$&$675$&$-1600$ \\
 \hline
  \phantom{$\Big|$}$\left[\Sigma_{+1}\Sigma_{0}\Sigma_{+1}\bfd_{-2/3}\right]^{\phantom{l}}$ & 
& \\ 
\phantom{$\big|$}$O^\uparrow(\epsilon^3;\se_{+1},\se_{+1},\for_{0})$ & $4725$&$800$ \\
\phantom{$\big|$}$O^\downarrow(\epsilon^3;\se_{+1},\se_{+1},\for_{0})$&$675$&$-1600$ \\
 \hline
  \phantom{$\Big|$}$\left[\Sigma_{+1}\Sigma_{+1}\Sigma_{0}\bfd_{-2/3}\right]^{\phantom{l}}$ & 
& \\ 
\phantom{$\big|$}$O^\uparrow(\epsilon^3;\se_{+1},\for_0,\for_{0})$&$4275$&$1200$ \\
\phantom{$\big|$}$O^\downarrow(\epsilon^3;\se_{+1},\for_0,\for_{0})$&$1575$&$-400 $ \\
\phantom{$\big|$}$O^\uparrow(\epsilon^3;\se_{+1},\se_0,\for_{0})$&$4275$&$800$ \\
\phantom{$\big|$}$O^\downarrow(\epsilon^3;\se_{+1},\se_0,\for_{0})$&$675$&$-1600$ \\
\hline 
\phantom{$\Big|$}$\sum_\mathbf{R}O(\epsilon^3;\mathbf{R})$        &  20925  & 3400 \\
\hline
\end{tabular}
   \caption{Possible ${\cal O}(\epsilon^3)$ corrections  to $Y^D_{22}$. }  
  \label{tab:Yld22eps3}
\end{table}
As regards the mass ratio $m_s/m_\mu$ we see from table~\ref{tab:Yld22eps2}
that a rather large value is obtained.  However, by inspecting the set of
${\cal O}(\epsilon^3)$ corrections we have found that they can add up to
produce a surprisingly large coefficient $20925/(\sqrt{60})^3\sim 45$ (see
table~\ref{tab:Yld22eps3}).  This can increase $m_\mu$ by 50\% and bring the
value of $3m_s/m_\mu$ quite close to the range given in eq.~(\ref{eq:GJ}).
This gives one example of a case when a non-hierarchical coefficient, rather
than being an ${\cal O}(1)$ number, is large enough to compensate for one
additional factor of $\epsilon$.
\vspace{-.3cm}
\subsection{The $Y^D_{11}$ entry and the down and electron masses}
\vspace{-2mm} 
The contributions to the down-quark and electron masses are listed in
table~\ref{tab:Yld11eps3}.  We see that all the four possibilities involve a
$\bforty$ and, as was mentioned above, because of the $SU(5)$ indices
properties of the $\forty$ at this order the electron mass vanishes.
$Y^\ell_{11}$ is thus promoted to ${\cal O}(\epsilon^4)$ and this also implies
${\rm det}\, Y^\ell \sim \epsilon^7$ instead than the naive estimate
eq.~(\ref{eq:det1}).  It is interesting that in the present scheme the FN
mechanism is not only able to split the $SU(5)$ mass degeneracies by means of
non hierarchical coefficients, but it can also produce a relative hierarchy
between the lepton and down-type quark masses of the same generation.  
 As regards the second GUT relation in eq.~(\ref{eq:GJ}), it can be reproduced
if, for example, the ${\cal O}(\epsilon^4)$ correction to the down-quark
Yukawa coupling remains small, while the coefficient for the electron is
$\approx 15$, and we have just seen that numbers of this size are certainly
possible.
 \begin{table}[htbp] 
\begin{tabular}{|l|r|r|}
\hline
 \phantom{$\Big|^l$} &$\frac{Y_{11}^\ell}{(\epsilon/\sqrt{60})^3}$ & $\frac{Y_{11}^d}{(\epsilon/\sqrt{60})^3}$ \\
\hline
\phantom{$\Big|$}$\left[\Sigma_{+1}\bfd_{-2/3}\Sigma_{+1}\Sigma_{+1}\right]^{\phantom{l}}$ & 
& \\ 
\phantom{$\big|$}$O(\epsilon^3;\fv_{+3},\t_{+11/3},\bforty_{+8/3})$ & $0$&$100$ \\
\phantom{$\big|$}$O(\epsilon^3;\for_{+3},\t_{+11/3},\bforty_{+8/3})$ & $0$&$500$ \\
%
\hline
 \phantom{$\Big|$}$\left[\bfd_{-2/3}\Sigma_{+1}\Sigma_{+1}\Sigma_{+1}\right]^{\phantom{l}}$ & 
& \\ 
\phantom{$\big|$}$O(\epsilon^3;\t_{+14/3},\t_{+11/3},\bforty_{+8/3})$ & $0$&$50$ \\
\phantom{$\big|$}$O(\epsilon^3;\fif_{+14/3},\t_{+11/3},\bforty_{+8/3})$ & $0$&$250$ \\
%
\hline 
       \phantom{$\Big|$}$\sum_\mathbf{R}O(\epsilon^3;\mathbf{R})$                              &  0  & 900 \\
\hline
\end{tabular}
   \caption{Operators contributing  to $Y^D_{11}$ at ${\cal O}(\epsilon^3)$. }  
  \label{tab:Yld11eps3}
\end{table}
\vspace{-.3cm}
\subsection{Other ${\cal O}(\epsilon^2)$ and ${\cal O}(\epsilon^3)$ entries:
  $Y^D_{32}$, $Y^D_{31}$ and $Y^D_{23}$.}
\vspace{-2mm}                                
For completeness, we list in tables~\ref{tab:Yld32eps2}, \ref{tab:Yld31eps3}
and \ref{tab:Yld23eps3} the coefficients of the operators contributing
respectively to $Y^D_{32}$ at ${\cal O}(\epsilon^2)$ and to  
$Y^D_{31}$ and $Y^D_{23}$ at ${\cal O}(\epsilon^3)$.  The last two entries
$Y^D_{12}$ and $Y^D_{13}$ are highly suppressed (at least as $\epsilon^4$ and
$\epsilon^5$) and we have not computed them. In any case they give only
negligible corrections to mass ratios and mixing angles.
%
\begin{table}[h!!] 
\begin{tabular}{|l|r|r|}
\hline
 \phantom{$\Big|^l$} &$\frac{Y_{32}^\ell}{(\epsilon/\sqrt{60})^2}$ & $\frac{Y_{32}^d}{(\epsilon/\sqrt{60})^2}$ \\
\hline
\phantom{$\Big|$}$\left[\bfd_{-2/3}\Sigma_{-1}\Sigma_{-1}\right]^{\phantom{l}}$ & 
& \\ 
\phantom{$\big|$}$O(\epsilon^2;\t_{-4/3},\t_{-1/3})$ & $-36$&$-1$ \\ [-.2pt]
\phantom{$\big|$}$O(\epsilon^2;\t_{-4/3},\fif_{-1/3})$ & $0$&$-25$ \\ [-.2pt]
\phantom{$\big|$}$O(\epsilon^2;\t_{-4/3},\bforty_{-1/3})$ & $0$&$-50$ \\ [-.2pt]
 \phantom{$\big|$}$O(\epsilon^2;\fif_{-4/3},\t_{-1/3})$ & $0$&$-5$ \\ [-.2pt]
\phantom{$\big|$}$O(\epsilon^2;\fif_{-4/3},\fif_{-1/3})$ & $0$&$5$ \\
\hline
\phantom{$\Big|$}$\left[\Sigma_{-1}\bfd_{-2/3}\Sigma_{-1}\right]^{\phantom{l}}$ & 
& \\ 
\phantom{$\big|$}$O(\epsilon^2;\fv_{-1},\t_{-1/3})$ & $18$&$-2$ \\ [-.2pt]
\phantom{$\big|$}$O(\epsilon^2;\fv_{-1},\fif_{-1/3})$ & $0$&$-10$ \\ [-.2pt]
\phantom{$\big|$}$O(\epsilon^2;\for_{-1},\t_{-1/3})$ & $-90$&$-10$ \\ [-.2pt]
\phantom{$\big|$}$O(\epsilon^2;\for_{-1},\bforty_{-1/3})$ & $0$&$-200$ \\ [-.2pt]
\phantom{$\big|$}$O(\epsilon^2;\se_{-1},\fif_{-1/3})$ & $0$&$100$ \\
\hline
\phantom{$\Big|$}$\left[\Sigma_{-1}\Sigma_{-1}\bfd_{-2/3}\right]^{\phantom{l}}$ & 
& \\ 
\phantom{$\big|$}$O(\epsilon^2;\fv_{-1},\for_{0})$ & $45$&$-20$ \\ [-.2pt]
\phantom{$\big|$}$O^\uparrow(\epsilon^2;\for_{-1},\for_{0})$ & $285$&$-60$ \\ [-.2pt]
\phantom{$\big|$}$O^\downarrow(\epsilon^2;\for_{-1},\for_{0})$ & $105$&$20$ \\ [-.2pt]
\phantom{$\big|$}$O(\epsilon^2;\se_{-1},\for_{0})$ & $225$&$-200$ \\
\hline 
 \phantom{$\Big|$}$\sum_\mathbf{R}O(\epsilon^2;\mathbf{R})$ 
&  $552$  & $-458$ \\
\hline
\end{tabular}
   \caption{Operators contributing  to $Y^D_{32}$ at ${\cal O}(\epsilon^2)$. 
}  
  \label{tab:Yld32eps2}
\end{table}
%
%
\begin{table}[h!!] 
\begin{tabular}{|l|r|r|}
\hline
 \phantom{$\Big|^l$} &$\frac{Y_{31}^\ell}{(\epsilon/\sqrt{60})^3}$ & $\frac{Y_{31}^d}{(\epsilon/\sqrt{60})^3}$ \\
\hline
\phantom{$\Big|$}$\left[\Sigma_{-1}\bfd_{-2/3}\Sigma_{-1}\Sigma_{-1}\right]^{\phantom{l}}$ & 
& \\ 
\phantom{$\big|$}$O(\epsilon^3;\fv_{-1},\t_{-1/3},\t_{+2/3})$ & $-108$&$2$ \\ [-.2pt]
\phantom{$\big|$}$O(\epsilon^3;\fv_{-1},\t_{-1/3},\fif_{+2/3})$ & $0$&$50$ \\ [-.2pt]
\phantom{$\big|$}$O(\epsilon^3;\fv_{-1},\t_{-1/3},\bforty_{+2/3})$ & $0$&$100$ \\ [-.2pt]
\phantom{$\big|$}$O(\epsilon^3;\fv_{-1},\fif_{-1/3},\t_{+2/3})$ & $0$&$10$ \\ [-.2pt]
\phantom{$\big|$}$O(\epsilon^3;\fv_{-1},\fif_{-1/3},\fif_{+2/3})$ & $0$&$-10$\\ [-.2pt]
\phantom{$\big|$}$O(\epsilon^3;\for_{-1},\t_{-1/3},\t_{+2/3})$ & $540$&$10$ \\[-.2pt]
\phantom{$\big|$}$O(\epsilon^3;\for_{-1},\t_{-1/3},\fif_{+2/3})$ & $0$&$250$ \\[-.2pt]
\phantom{$\big|$}$O(\epsilon^3;\for_{-1},\t_{-1/3},\bforty_{+2/3})$ & $0$&$500$ \\[-.2pt]
\phantom{$\big|$}$O(\epsilon^3;\for_{-1},\bforty_{-1/3},\t_{+2/3})$ &$0$&$-200$ \\[-.2pt]
\phantom{$\big|$}$O^\uparrow(\epsilon^3;\for_{-1},\bforty_{-1/3},\bforty_{+2/3})$ &$0$&$800$ \\[-.2pt]
\phantom{$\big|$}$O^\downarrow(\epsilon^3;\for_{-1},\bforty_{-1/3},\bforty_{+2/3})$ & $0$&$-1400$ \\[-.2pt]
\phantom{$\big|$}$O(\epsilon^3;\se_{-1},\fif_{-1/3},\t_{+2/3})$ & $0$&$-100$ \\[-.2pt]
\phantom{$\big|$}$O(\epsilon^3;\se_{-1},\fif_{-1/3},\fif_{+2/3})$ & $0$&$100$ \\
\hline
 \phantom{$\Big|$}$\left[\Sigma_{-1}\Sigma_{-1}\bfd_{-2/3}\Sigma_{-1}\right]^{\phantom{l}}$ & 
& \\ 
\phantom{$\big|$}$O(\epsilon^3;\fv_{-1},\for_{0},\t_{+2/3})$ & $-270$&$20$ \\[-.2pt]
\phantom{$\big|$}$O(\epsilon^3;\fv_{-1},\for_{0},\bforty_{+2/3})$ & $0$&$400$ \\[-.2pt]
\phantom{$\big|$}$O(\epsilon^3;\fv_{-1},\se_{0},\fif_{+2/3})$ & $0$&$-200$ \\[-.2pt]
\phantom{$\big|$}$O^\uparrow(\epsilon^3;\for_{-1},\for_{0},\t_{+2/3})$ & $-1710$&$60$ \\[-.2pt]
\phantom{$\big|$}$O^\downarrow(\epsilon^3;\for_{-1},\for_{0},\t_{+2/3})$ &$-630$&$-20$ \\[-.2pt]
\phantom{$\big|$}$O^\uparrow(\epsilon^3;\for_{-1},\for_{0},\bforty_{+2/3})$ & $0$&$-1200$ \\[-.2pt]
\phantom{$\big|$}$O^\downarrow(\epsilon^3;\for_{-1},\for_{0},\bforty_{+2/3})$ & $0$&$400$ \\[-.2pt]
\phantom{$\big|$}$O(\epsilon^3;\for_{-1},\exx{$\bfifty_{0}$},\bforty_{+2/3})$ & $-$&$-$ \\[-.2pt]
\phantom{$\big|$}$O(\epsilon^3;\for_{-1},\se_{0},\fif_{+2/3})$ & $0$&$1000$ \\[-.2pt]
\phantom{$\big|$}$O(\epsilon^3;\se_{-1},\for_{0},\t_{+2/3})$ & $-1350$&$200$ \\[-.2pt]
\phantom{$\big|$}$O(\epsilon^3;\se_{-1},\for_{0},\bforty_{+2/3})$ & $0$&$4000$ \\[-.2pt]
\phantom{$\big|$}$O^\uparrow(\epsilon^3;\se_{-1},\se_{0},\fif_{+2/3})$ & $0$&$-400$ \\[-.2pt]
\phantom{$\big|$}$O^\downarrow(\epsilon^3;\se_{-1},\se_{0},\fif_{+2/3})$ & $0$&$800$ \\

\hline 
 \phantom{$\Big|$}$\sum_\mathbf{R}O(\epsilon^3;\mathbf{R})$ &  $-3528$  & $5172$ \\
\hline
\end{tabular}
   \caption{Operators contributing  to $Y^D_{31}$ at ${\cal O}(\epsilon^3)$. 
}  
  \label{tab:Yld31eps3}
\end{table}
%
%
\begin{table}[h!!] 
\begin{tabular}{|l|r|r|}
\hline
 \phantom{$\Big|^l$} &$\frac{Y_{23}^\ell}{(\epsilon/\sqrt{60})^3}$ & $\frac{Y_{23}^d}{(\epsilon/\sqrt{60})^3}$ \\
\hline
\phantom{$\Big|$}$\left[\Sigma_{+1}\Sigma_{+1}\bfd_{-2/3}\Sigma_{+1}\right]^{\phantom{l}}$ & 
& \\ 
\phantom{$\big|$}$O(\epsilon^3;\se_{+1},\for_{0},\t_{+2/3})$ & $-1350$&$200$ \\
\phantom{$\big|$}$O(\epsilon^3;\se_{+1},\for_{0},\bforty_{+2/3})$ & $0$&$4000$ \\
\phantom{$\big|$}$O^\uparrow(\epsilon^3;\se_{+1},\se_{0},\fif_{+2/3})$ & $0$&$-400$ \\
\phantom{$\big|$}$O^\downarrow(\epsilon^3;\se_{+1},\se_{0},\fif_{+2/3})$ &$0$&$800$ \\
\hline
\phantom{$\Big|$}$\left[\Sigma_{+1}\Sigma_{+1}\Sigma_{+1}\bfd_{-2/3}\right]^{\phantom{l}}$ & 
& \\ 
\phantom{$\big|$}$O(\epsilon^3;\se_{+1},\for_{0},\fv_{-1})$ & $225$&$-200$ \\
\phantom{$\big|$}$O^\uparrow(\epsilon^3;\se_{+1},\for_{0},\for_{-1})$ & $4275$&$1200$ \\
\phantom{$\big|$}$O^\downarrow(\epsilon^3;\se_{+1},\for_{0},\for_{-1})$ & $1575$&$-400$ \\
\phantom{$\big|$}$O^\uparrow(\epsilon^3;\se_{+1},\se_{0},\fv_{-1})$ & $-4725$&$800$ \\
\phantom{$\big|$}$O^\downarrow(\epsilon^3;\se_{+1},\se_{0},\fv_{-1})$ & $-675$&$-1600$ \\
\phantom{$\big|$}$O^\uparrow(\epsilon^3;\se_{+1},\se_{0},\for_{-1})$ & $4725$&$800$ \\
\phantom{$\big|$}$O^\downarrow(\epsilon^3;\se_{+1},\se_{0},\for_{-1})$ & $675$&$-1600$ \\
\hline 
 \phantom{$\Big|$}$\sum_\mathbf{R}O(\epsilon^3;\mathbf{R})$ 
&  $4724$  & $3600$ \\
\hline
\end{tabular}
   \caption{Operators contributing to $Y^D_{23}$ at ${\cal O}(\epsilon^3)$ . 
}  
  \label{tab:Yld23eps3}
\end{table}

\section{Discussion}
\vspace{-2mm} 
Before discussing what can be learned from our results, let us resume briefly
the main steps of the whole procedure.  We have selected a set of $U(1)_F$
charges suitable to reproduce the observed fermion mass hierarchy
eqs.~(\ref{eq:ratiosdl}) and (\ref{eq:ratiosu}), and satisfying our
theoretical prejudice that each fermion multiplet should be univocally
identified by the GUT-flavor symmetry.  By assuming a common mass for the
heavy states and universality for the fundamental Yukawa couplings we have
reduced the number of free parameters to one: the dimensionless symmetry
breaking parameter $\epsilon$. We have then computed the effective down-quarks
and lepton Yukawa matrices by including at each order in $\epsilon$ all the
possible operators, except for a few cases when eliminating some contribution
was mandatory (this was done consistently, by assuming that FN fields in
specific $SU(5)\times U(1)_F$ representations are absent).  We have seen that
at leading order $b$-$\tau$ unification is preserved, while at order
$\epsilon^2$ and higher the lepton and down-quark Yukawa matrices differ, and
not only in the non-hierarchical coefficients, but possibly also in the order
of their hierarchical suppression.

The lepton Yukawa matrix $Y^\ell$ that we have obtained is not particularly
predictive.  This is because the ratio $m_\mu/m_\tau$ has been fitted to
determine the value of $\epsilon$, $m_e$ got promoted to ${\cal
  O}(\epsilon^4)$ and, since we have limited our analysis to ${\cal
  O}(\epsilon^3)$, has not been computed, and quantitative results for the
leptonic mixing angles require including a model for neutrino masses. This
implies more structure and additional assumptions, and goes beyond the scope
of this study.  The down-quarks Yukawa matrix $Y^d$ is more informative.
Numerically we obtain
\begin{equation}
\label{eq:Ydnum}
Y^d \approx 
\begin{pmatrix}
  1.9\,\epsilon^3  &  \sim \epsilon^5  &  \sim \epsilon^4  \\[-0pt]
  9.9\,\epsilon^3  & -3.3\,\epsilon^2  &  7.8\,\epsilon^3  \\[-0pt]
 11.1\,\epsilon^3  & -7.6\,\epsilon^2  &  2.3\,\epsilon   
\end{pmatrix},
\end{equation}
where $\epsilon \sim 0.037$. From $Y^d$  we obtain the mass ratios
\begin{equation}
\label{eq:ratiosdnum}
\frac{m_s}{m_b}\approx 0.05,\qquad \frac{m_d}{m_s}\approx 0.02,
\end{equation}
together with the down-quarks L-handed  mixing matrix
\begin{equation}
\label{eq:mixnum}
 V^d_L \approx 
\begin{pmatrix}
 \ 0.99  & \ 0.11  &  0.007 \\[-0pt]
 \ 0.11  & -0.98   &  -0.12   \\[-0pt]
  0.006  & -0.12   & \  0.99   
\end{pmatrix} .
\end{equation}
The ratios in eq.~(\ref{eq:ratiosdnum}) suggest that $m_s$ is about a factor
of 2 too large (experimentally $m_s/m_b \sim 0.01$-$0.02$, $m_d/m_s \sim
0.04$-$0.06$). This is also suggested by the GUT relation $3 m_s/m_\mu$ whose
central value in eq.~(\ref{eq:GJ}) would be  reproduced rather precisely 
if $m_s$ were half its size.  As regards the mixing matrix $V^d_L$, it has a
quite reasonable structure: if the corresponding matrix in the up-quark
sector has a similar structure, it is likely that the 
CKM matrix could be correctly reproduced.  Of course, one could improve the
numerical performance of the model by inspecting carefully
tables~\ref{table:Yd33} to~\ref{tab:Yld23eps3} and eliminating (consistently)
specific contributions.  However, in our opinion there is not much to learn
from the construction of an {\it ad hoc} realization, even if quantitatively
successful.  For example, it would not be surprising if starting from the
second set of charges in table~\ref{table:2}, and with a careful choice of
the relevant contributions, one could also obtain acceptable results.  Also,
other charge assignments different from the ones given in table~\ref{table:2} could
be viable since, as we have learned, starting from a set of charges that yields a
(naive) hierarchy milder than the one observed, it can still be possible to
generate the correct hierarchical pattern eqs.~(\ref{eq:ratiosdl})
and~(\ref{eq:ratiosu}).

Instead, we think that something more interesting can be learned by
considering some general features of the model.  The Abelian flavor symmetry
was introduced to generate a hierarchy between the entries of the Yukawa
matrices. While it is generally believed that from the observed hierarchy it
should be possible to reconstruct the Abelian charges, we have shown that in
some cases there is no direct relation between the charges and the
hierarchical suppression.  As regards the non-hierarchical coefficients, they
are ultimately determined by the $SU(5)$ symmetry.  However, a glance at $Y^d$
in eq.~(\ref{eq:Ydnum}) shows that we should not expect to observe any clear
trace of this symmetry in experimentally measurable quantities.  This is
because the number of $SU(5)$ coefficients contributing to $Y^U$, $Y^d$ and
$Y^\ell$ is much larger than the number of entries, and in turn the number of
entries is much larger than the number of observables.  It is then conceivable
that the unsuccess in trying to understand the origin of fermion masses could
be due to the very nature of a problem in which the amount of physically
accessible information is not sufficient to identify the solution.  In our
example, in spite of the fact that there is only one free parameter and that
everything else is computable, identifying the simple $SU(5)\times U(1)_F$
symmetry could well remain out of the reach of theoretical efforts. However,
if in the future more precise measurements will confirm with high precision
some of the observed regularities, and if new regularities will emerge, this
would be a convincing hint that only a few fundamental parameters concur to
determine the fermion mass spectrum, and would disprove schemes like the one
we have discussed.

 \vspace{-.4cm}
 \begin{acknowledgments}
{} \vspace{-.3cm}
  The idea of breaking the Abelian flavor symmetry with the adjoint of $SU(5)$
  was suggested long ago to one of us (E.N.)  by Z. Berezhiani.  We
  acknowledge conversations with J. Mira, W. Ponce, D. Restrepo and W.
  Tangarife.
 \end{acknowledgments}

 \vspace{-.5cm}

\appendix  
 \section{Group theory}
\label{sec:group}
\vspace{-0mm}
\subsection{Tensor products}
\vspace{-0mm}
We list some useful tensor products involving the $\bfd$ and the
$\mathbf{24}$-dimensional adjoint $\Sigma$ containing the Higgs fields (in
some cases our conventions for the conjugate representations differ from the
ones used in \cite{Slansky:1981yr}.)
%
\begin{eqnarray} \nonumber
\mathbf{\overline{5} \otimes \overline{5}} \!\!&=&\!\! \mathbf{ \overline{10}  \oplus \overline{15}} \\ [-1pt] \nonumber
\mathbf{10 \otimes \overline{5}} \!\!&=&\!\! \mathbf{5 \oplus 45} \\  [-1pt]\nonumber
\mathbf{\overline{10} \otimes \overline{5}} \!\!&=&\!\! \mathbf{ 10 \oplus \overline{40}} \\ [-1pt] \nonumber
\mathbf{15 \otimes \overline{5}} \!\!&=&\!\! \mathbf{5 \oplus 70} \\ [-1pt] \nonumber
\mathbf{\overline{15} \otimes \overline{5}} \!\!&=&\!\! \mathbf{ \overline{35} \oplus \overline{40}} \\ [-1pt] \nonumber
\mathbf{40 \otimes \overline{5}} \!\!&=&\!\! \mathbf{ 10 \oplus 15\oplus 175} \\ [-1pt] \nonumber
\mathbf{\overline{40} \otimes \overline{5}} \!\!&=&\!\! \mathbf{45 \oplus
  \overline{50}\oplus \overline{105}} \\ [-1pt] 
\label{eq:tensor} 
\mathbf{\overline{45} \otimes \overline{5}} \!\!&=&\!\! \mathbf{ \overline{10}\oplus 40\oplus\overline{175}} \\ [-1pt] \nonumber
\mathbf{\overline{70} \otimes \overline{5}} \!\!&=&\!\! \mathbf{ \overline{15} \oplus \overline{160} \oplus \overline{175}} \\ [-1pt] \nonumber
\mathbf{5\otimes 24  } \!\!&=&\!\! \mathbf{5 \oplus 45  \oplus 70}\\ [-1pt] \nonumber
\mathbf{10\otimes 24} \!\!&=&\!\! \mathbf{ 10 \oplus 15 \oplus \overline{40} \oplus 175} \\ [-1pt] \nonumber 
\mathbf{15\otimes 24} \!\!&=&\!\! \mathbf{ 10 \oplus 15 \oplus 160 \oplus 175} \\ [-1pt] \nonumber
\mathbf{40\otimes 24} \!\!&=&\!\! \mathbf{\overline{10} \oplus {35} \oplus 40\oplus 40\oplus \overline{175}\oplus  
\overline{210}\oplus 450^\prime }\\ [-1pt] \nonumber
\mathbf{ {45}\otimes{24}  } \!\!&=&\!\!  \mathbf{ 5 \oplus 45\oplus 45 \oplus  \overline{50}  
\oplus 70  \oplus \overline{105}  \oplus 280 \oplus \overline{480} } \\ [-1pt] \nonumber
\mathbf{ {70}\otimes {24}  } 
 \!\!&=&\!\! \mathbf{ 5 \oplus {45}\oplus  {70}\oplus{70}  
  \oplus {280}  \oplus  {280'}   \oplus \overline{450}  \oplus \overline{480}.} 
\vspace{-10pt}
\end{eqnarray}
\subsection{Vertices}
\vspace{-0mm}
The fundamental vertices involve $\bfd_a$ and the adjoint $\Sigma^a_b$.  They
have the general form $-i \lambda {\cal V}$ where $\lambda$ is assumed
universal and $ {\cal V}=\mathbf{R}\bfd\mathbf{R'} $ or $\mathbf{R}\Sigma\mathbf{R'}$, with
$\mathbf{R},\mathbf{R'}=\fv,\,\t,\,\fif,\,\for,\dots$.  
The relevant field contractions ${\cal V}$ including their symmetry factors are:
\begin{eqnarray} 
&&  \bfd_a \bfv_b \t^{ba}  \qquad \quad\  \bfd_a \bfv_b \fif^{ba} \hspace{3.5cm} \\
&& \frac{1}{2}\, \bfd_a \bfor_{bc}^a \t^{cb} \quad\ \frac{1}{2}\,\bfd_a \bse_{bc}^a \fif^{cb} \\ 
&&\frac{1}{2}\bfv_a \bfor_{bc}^n\mathbf{\bar{40}}_{nqr}\epsilon^{abcqr} =
-\frac{1}{4}\bfv_a \bfor_{bc}^n\mathbf{\bar{40}}_{qrn}\epsilon^{abcqr}       \\
&&  \bfv_a \Sigma^a_b \fv^b \qquad    \quad  \quad\  \  
\bfv_a \Sigma^c_b \for^{ba}_c \quad \   \quad\ 
\bfv_a \Sigma^c_b \se^{ba}_c \\
&& \bt_{ab} \Sigma^b_c \t^{ca} \quad  \quad\ \ 
\fif_{ab} \Sigma^b_c \fif^{ca} \quad\ \ 
\bfif_{ab} \Sigma^b_c \t^{ca} \\
 &&
\label{v40}
 \mathbf{40}^{abc}{\Sigma^\uparrow}_c^d\mathbf{\bar{40}}_{dba}
 \quad    \ 
 \mathbf{40}^{abc}{\Sigma^\downarrow}_a^d\mathbf{\bar{40}}_{dbc} \\ 
\label{v45}
&& \bfor_{ab}^c{\Sigma^\uparrow}^b_d\for^{da}_c \quad\    \quad           
   \frac{1}{2}\, \bfor_{ab}^c{\Sigma^\downarrow}_c^d\for^{ba}_d  \\
\label{v70} 
&& \bse_{ab}^c{\Sigma^\uparrow}^b_d\se^{da}_c   \quad\   \quad   
   \frac{1}{2}\, \bse_{ab}^c{\Sigma^\downarrow}_c^d\se^{ba}_d     \\
&& \frac{1}{2} \mathbf{40}^{abc}\Sigma_b^d\t^{fg}\epsilon_{acdfg} =  
 \frac{1}{4} \mathbf{40}^{abc}\Sigma_c^d\t^{fg}\epsilon_{abdfg}    \\
&& \bfor_{ab}^c\Sigma^b_d\se^{da}_c.  
\end{eqnarray}
There are two inequivalent way of contracting the indices for the vertices
involving the $\Sigma$ with pairs of $\forty$, $\for$ and $\se$. They are
distinguished in eqs.~(\ref{v40}),(\ref{v45}) and (\ref{v70}) by an up-
($\Sigma^\uparrow$) or down-arrow ($\Sigma^\downarrow$) label.  This can be traced back to
the fact that these representations are contained twice in their tensor 
products with the adjoint (see the last three lines in (\ref{eq:tensor})).
At order higher than $\epsilon^3$ other representations and  other vertices can appear, like e.g. 
\begin{equation}
\frac{1}{2}\,\mathbf{{35}}^{abc} \Sigma^d_c \mathbf{\bar{35}}_{dab},  
\quad \ 
\mathbf{{40}}^{abc} \Sigma^d_b \mathbf{\bar{35}}_{dac}, \quad\  \mbox{etc}\dots\qquad 
\end{equation}
%

\subsection{Pointlike propagators}
\vspace{-0mm}
The pointlike propagators in momentum space needed to build the effective
operators are defined as $({-i}/{M})\,{\cal S}^{abc\dots}_{lmn\dots}$ where
${\cal S}$ denotes the index structure appropriate for the given FN
representations.  The ${\cal S}$-factors for FN fields in the $\fv$ and $\t$
can be derived with standard path integral methods, and are given in
eqs.~(\ref{five}) and~(\ref{ten}). The $\fif$ is obtained by symmetrizing the
$\t$ over $SU(5)$ indices, yielding eq.~(\ref{fifteen}).  The sum of $\t$ and
$\fif$ corresponds to the reducible $\mathbf{25}^{(r)}$ and is given in
eq.~(\ref{twentyfive}).  The tensor product $\t^{ab}\otimes \bfv_c=[\for\oplus \fv]^{ab}_c$
corresponds to the reducible ${\mathbf{50}^{(r)}}^{\,ab}_c$ 
antisymmetric in the two upper indices given in eq.~(\ref{fifty}). The irreducible fragment $\for^{ab}_c$
can be identified by singling out the `trace' part
${\mathbf{50}^{(r)}}^{\,ab}_a$ that corresponds to the irreducible $\fv^b$
fragment. Requiring $\for^{ab}_a=\for^{ab}_b=0$ we obtain:
\begin{eqnarray} \nonumber 
 {[}\for^{ab}_c \, \bfor_{lm}^n {]} \to  -4\, \delta^n_c\,\left[\delta^a_l\,\delta^b_m -
  \delta^a_m\,\delta^b_l\right]_{\mathbf{50}^{(r)}} - \qquad\quad && \\  
 \quad \left[ \delta^a_c\,(\delta^b_l\,\delta^n_m -
\delta^b_m\,\delta^n_l) - \delta^b_c\,(\delta^a_l\,\delta^n_m -
\delta^a_m\,\delta^n_l)\right]_{\fv}.  \quad  && 
\label{fortyfive1} 
\end{eqnarray}
The first term on the r.h.s corresponds to the $\mathbf{50}^{(r)}$ and the
second term is the $\fv$ piece.  The overall normalization is fixed by
requiring that the $\fv$ piece gives the same contribution to the operator
$\bfv\,\t\,\bfd$ than eq.~(\ref{five}).  By means of the identity
\begin{eqnarray}\nonumber
&&\frac{1}{2!} \epsilon^{abnij}\,\epsilon_{clmij}=
 \delta^a_c\,(\delta^b_l\,\delta^n_m - \delta^b_m\,\delta^n_l)\hspace{2.5cm} \\
&&\qquad - \delta^b_c\,(\delta^a_l\,\delta^n_m -
\delta^a_m\,\delta^n_l) +\delta^n_c\,(\delta^a_l\,\delta^b_m - \delta^a_m\,\delta^b_l),
\end{eqnarray}
(\ref{fortyfive1}) can be conveniently rewritten as in eq.~(\ref{fortyfive}).
The $\se$ is constructed in a similar way. It is contained in the tensor
product $\fif^{ab}\otimes \bfv_c= [\se\oplus \fv]^{ab}_c $ that corresponds to
the reducible ${\mathbf{75}^{(r)}}^{\,ab}_c$ symmetric in the two upper
indices. By imposing the `traceless' condition $\se^{ab}_a= \se^{ab}_b=0$ we
obtain
\begin{eqnarray} \nonumber 
{[}\se^{ab}_c \, \bse_{lm}^n {]} \to 6\,
\delta^n_c\,\left[\delta^a_l\,\delta^b_m +
\delta^a_m\,\delta^b_l\right]_{\mathbf{75}^{(r)}} - \qquad \qquad  \ && \\
\label{seventy1} 
\left[ \delta^a_c\,(\delta^b_l\,\delta^n_m +
  \delta^b_m\,\delta^n_l) + \delta^b_c\,(\delta^a_l\,\delta^n_m +
  \delta^a_m\,\delta^n_l)\right]_{\mathbf{5}}\,.\qquad &&
\end{eqnarray}
To fix the normalization one has to go to ${\cal O}(\epsilon^2)$ and compute
e.g. the entry (11) in table~\ref{table:eps2}.  

The $\mathbf{40}$ is contained in the tensor product $\t^{ab}\otimes \fv^c=
[\bt\oplus \mathbf{40}]^{abc}
$  that corresponds to
a reducible ${\mathbf{50'}^{(r)}}^{\,abc}$ antisymmetric in the first two 
indices. The three-index $\bt^{abc}$ fragment that we need to subtract 
in order to single out the $\forty$ is related 
to the two-index $\bt_{ij}$ through the conjugate of the following dual relations: 
\begin{equation}
\t_{abc} = \frac{1}{2!}\epsilon_{abc ij}\t^{ij}, \qquad
\t^{ij} = \frac{1}{3!}\epsilon^{ijabc}\t_{abc}. 
\end{equation}
The dual representations satisfy the identity   
\begin{equation}
\label{eq:tenrelation}
\frac{1}{3!} \t_{lmn}\bt^{abc}\epsilon_{abcij}=
\frac{1}{2!} \epsilon_{lmndf} \t^{df}\bt_{ij}=\epsilon_{lmnij}
\end{equation}
where in the last step eq.~(\ref{ten}) has been used.
This implies
\begin{equation}
\left[\t_{lmn}\,\bt^{abc}\right]\to 
\frac{1}{2!}\,
\epsilon_{pq lmn}
\epsilon^{pq abc}, 
\end{equation}
as can be easily checked by substituting this result in (\ref{eq:tenrelation})
and by using {$\epsilon^{pqabc} \epsilon_{abcij}=3!\,
  (\delta^p_i\delta^q_j-\delta^p_j\delta^q_i)$.  The expression for the
  $\mathbf{40}$ can be now obtained by subtracting the contribution of the
  $\bt$ from the $\mathbf{50^\prime}^{(r)}$:
\begin{equation}
[\mathbf{40}^{abc} \mathbf{\bar{40}}_{lmn}] 
 \to   
3(\delta^a_l\delta ^b_m - \delta^a_m\delta^b_l)\delta^c_n 
- \frac{1}{2}\epsilon^{ijabc}\epsilon_{ijlmn}.   
\end{equation}
This expression satisfies antisymmetry in the first two indices $
\mathbf{40}^{abc}=-\mathbf{40}^{bac}$ plus the 10 conditions
$\epsilon_{ijabc}\mathbf{40}^{abc}=0$ that can be used to fix the factor of 3
for the $\mathbf{50^\prime}^{(r)}$. Note that the last 10 conditions imply
that the $\forty$ does not contribute to the lepton mass operators.  This can
be understood by considering the vertex
$\forty^{abc}\,\Sigma^d_b\,\t^{fg}\epsilon_{acdfg}$ eq.~(\ref{v45}).  When the
$\t^{fg}$ is projected on the leptons ($f,g=4,5$), $\langle \Sigma\rangle $
gets restricted to the upper-left $3\times 3$ corner ($b,d=1,2,3$) that is
proportional to the identity $\delta^d_b$. Then the vertex collapses into the
10 vanishing conditions.  

The irreducible $\fifty$ (with Young tableau
\begin{picture}(12,10)(0,1)
\put(0,0){\framebox(5,5)}\put(5.5,0){\framebox(5,5)}
\put(0,5.5){\framebox(5,5)}\put(5.5,5.5){\framebox(5,5)}
\end{picture})
is a four index representation that appears at ${\cal O}(\epsilon^3)$ but only
in one case (the entry (38) in table~\ref{table:eps3b}). Constructing its
index structure and normalization is rather awkward so we have omitted the
$\fifty$ from our analysis.

At ${\cal O}(\epsilon^4)$ the $\mathbf{35}$ can appear.  Even if our analysis
is restricted to ${\cal O}(\epsilon^3)$, we present the ${\cal S}$ structure
for the $\mathbf{35}$ since it can be derived rather easily.  The $\mathbf{35}$ is
contained in the tensor product $\fif^{ab}\otimes
\fv^b=[\mathbf{35}\oplus\forty]^{abc}$ and corresponds to the Young tableau
\begin{picture}(18,6)(0,-1)
\put(0,0){\framebox(5,5)}
\put(5.5,0){\framebox(5,5)}
\put(11,0){\framebox(5,5)}
\end{picture}.
Being totally symmetric, its index structure is straightforwardly constructed
(see eq.~(\ref{thirtyfive})).  The overall normalization is fixed by requiring
that the $\forty$ irreducible fragment contained in the two-index symmetric
reducible $\mathbf{75'}^{(r)}$ ${[}\mathbf{75'}^{(r)\,abc} \,
\mathbf{\bar{75'}}_{lmn}^{(r)} {]} \sim \left(\delta^a_l\,\delta^b_m +
  \delta^a_m\,\delta^b_l\right)\delta^c_n$ reproduces the results obtained
with eq.~(\ref{forty}).  The $\mathbf{40}$ in the two-index symmetric
representation can be singled out by imposing the vanishing of the symmetric
combinations $\mathbf{40}^{abc}+\mathbf{40}^{acb}+\mathbf{40}^{cba}=0$.
Taking into account the symmetry in the first two indices this can be
expressed as a cyclical relation and gives 35 conditions. We obtain
\begin{eqnarray} 
\nonumber &&\!\!\!\!\!\!\!\!\!   
{[}\forty^{abc} \, \bforty_{lmn} {]} \to  4\,
\left[\left(\delta^a_l\,\delta^b_m + 
  \delta^a_m\,\delta^b_l\right)\delta^c_n\right]_{\mathbf{75'}^{(r)}}  -  \\ \label{fortysym}
 && 2 \left[ \left(\delta^a_l\,\delta^c_m + 
  \delta^a_m\,\delta^c_l\right) \delta^b_n +
\left(\delta^c_l\,\delta^b_m + 
  \delta^c_m\,\delta^b_l\right) \delta^a_n 
\right]_{\mathbf{35}}. \qquad \ \ 
\end{eqnarray}
In summary, the relevant index structures that we have evaluated are:
\begin{eqnarray} 
\label{five} && \!\!\!\!\!
{[}\fv^a\,\bfv_b{]} \to\,  \delta^a_b  \\   
 \label{fortyfive}  &&\!\!\!\!\!
{[}\for^{ab}_c \, \bfor_{lm}^n{]} \to  
 -3\left(\delta^a_l\delta^b_m -
  \delta^a_m\delta^b_l\right)\delta^n_c
- \frac{1}{2}\epsilon^{abnij}\epsilon_{lmcij}\qquad\ \ \\  
\label{ten} &&\!\!\!\!\!
{[} {\mathbf{10}}^{ab}\, \mathbf{\bar {10}}_{lm} {]}  \to
\,  \left(\delta^a_l\, \delta^b_m- \delta^a_m\,
  \delta^b_l\right)  \\ [3pt]
\label{fifteen} &&\!\!\!\!\!
{[} {\mathbf{15}}^{ab}\, \mathbf{\bar {15}}_{lm} {]}  
\to\,  \left(\delta^a_l\, \delta^b_m+ \delta^a_m\,
  \delta^b_l\right)  \\ 
\nonumber  &&\!\!\!\!\!
{[}\tfv^{abc} \, \btfv_{lmn} {]} \to  2\,
\left[\left(\delta^a_l\,\delta^b_m + 
  \delta^a_m\,\delta^b_l\right)\, \delta^c_n + \right. \\ 
\label{thirtyfive} 
&& \qquad \left. \left(\delta^a_l\,\delta^c_m + 
  \delta^a_m\,\delta^c_l\right)\, \delta^b_n +
\left(\delta^c_l\,\delta^b_m +   \delta^c_m\,\delta^b_l\right)\, \delta^a_n 
\right]\, \\ 
&&\!\!\!\!\! \label{forty} 
{[}\mathbf{40}^{abc} \, \mathbf{\bar{40}}_{lmn} {]} \to 
3\left(\delta^a_l\delta^b_m 
-  \delta^a_m\delta^b_l\right)\delta_n^c- \frac{1}{2}\epsilon^{ijabc}
\epsilon_{ijlmn}\\ 
 &&\!\!\!\!\!
{[}\se^{ab}_c \, \bse_{lm}^n {]} \to 6
\left(\delta^a_l\,\delta^b_m +
  \delta^a_m\,\delta^b_l\right) \delta^n_c  \nonumber   \\ \label{seventy} 
&& \qquad
- \left(\delta^b_l\,\delta^n_m -
  \delta^b_m\,\delta^n_l\right)\delta^a_c + \left(\delta^a_l\,\delta^n_m +
  \delta^a_m\,\delta^n_l\right)\delta^b_c\,.  
\end{eqnarray}
The index structures for the reducible $\mathbf{25}^{(r)}$ and
$\mathbf{50}^{(r)}$ used in sec.~\ref{sec:btau}  are:
%
        \begin{eqnarray} 
          \label{twentyfive} &&\!\!\!\!\!
       {[} {\mathbf{25}^{(r)}}^{\,ab}\, \mathbf{\bar {25}}^{(r)}_{lm} {]}  \to\ \ 
       2\,  \delta^a_l\, \delta^b_m   \\ 
       \label{fifty} &&\!\!\!\!\!
       {[}\mathbf{50}^{(r)\,ab}_c \, \mathbf{\bar{50}}_{lm}^{(r)\,n} {]} \to - 4\,
       \delta^n_c\,\left[\delta^a_l\,\delta^b_m -
        \delta^a_m\,\delta^b_l\right].  \qquad   \qquad
       \end{eqnarray}

\bigskip 

\section{Tables of results} 
\label{sec:tables} 
\vspace{-0mm}
In tables~\ref{table:eps1} to~\ref{table:eps3b} we collect 
the coefficients of the mass operators contributing to the effective Yukawa
couplings $Y^d$ and $Y^\ell$ at ${\cal O}(\epsilon,\,\epsilon^2,\,\epsilon^3)$.
The operators are evaluated with a factor $-i{\cal V}$ for each vertex
and $-i{\cal S}/M$ for each propagator, and by dividing the result by~$i$. 
\begin{table}[htbp] 
\begin{center}
\begin{tabular}{|r|c|r|r|c|}
\hline
\phantom{$\Big|$} &${\cal O}(\epsilon)$ & $\frac{Y^\ell}{\epsilon/\sqrt{60}}$ & $\frac{Y^d}{\epsilon/\sqrt{60}}$ & \\ 
\hline
1)\phantom{$\Big|$}& $\mathbf{10}$ & $6$ & $1$ & $\lower 0pt\hbox{$\mathbf{\bar{\phi}_d \Sigma}$}$  \\  [-5pt]
2)\phantom{$\Big|$}& $\mathbf{15}$ & $0$ & $5$ & \\ [-0pt]
\hline
3)\phantom{$\Big|$}& $\mathbf{5}$ & $-3$ & $2$ &$\lower 0pt\hbox{$\mathbf{\Sigma \bar{\phi}_d}$}$ \\  [-5pt]
4)\phantom{$\Big|$} & $\mathbf{45}$ & $15$ & $10$ &  \\ [-0pt]
\hline
\end{tabular}
\caption{Operators contributing to $Y^\ell$ and $Y^d$ at ${\cal O}(\epsilon)$.
\protect \label{table:eps1}
}
\end{center}
\end{table}
%
%
\vspace{-0cm}
\begin{table}[htbp] 
\begin{center}
\begin{tabular}{|r|c|r|r|c|}
\hline
\phantom{$\Big|$} &${\cal O}(\epsilon^2)$ & $\frac{Y^\ell}{(\epsilon/\sqrt{60})^2}$   &  $\frac{Y^d}{(\epsilon/\sqrt{60})^2}$ & \\ 
\hline
1)\phantom{$\Big|$}& $\mathbf{10}$  $\mathbf{10}$ & $-36$ & $-1$ & $\lower 0pt\hbox{$\mathbf{\bar{\phi}_d \Sigma \Sigma}$}$  \\  [-5pt]
2)\phantom{$\Big|$}& $\mathbf{10}$ $\mathbf{15}$ & $0$ & $-25$ &  \\  [-5pt]
3)\phantom{$\Big|$}& $\mathbf{10}$ $\mathbf{\bar{40}}$ & $0$ & $-50$ &   \\  [-5pt]
4)\phantom{$\Big|$}& $\mathbf{15}$ $ \mathbf{10}$ & $0$ & $-5$ & \\ [-5pt]
5)\phantom{$\Big|$}& $\mathbf{15}$ $\mathbf{15}$ & $0$ & $5$ & \\ [-0pt]
\hline
6)\phantom{$\Big|$}& $\mathbf{5}$ $\mathbf{10}$ & $18$ & $-2$ & $\lower 0pt\hbox{$\mathbf{ \Sigma \bar{\phi}_d \Sigma}$}$  \\  [-5pt]
7)\phantom{$\Big|$}& $\mathbf{5}$ $\mathbf{15}$ & $0$ & $-10$ &  \\  [-5pt]
8)\phantom{$\Big|$}& $\mathbf{45}$ $\mathbf{10}$ & $-90$ & $-10$ &   \\  [-5pt]
9)\phantom{$\Big|$}& $\mathbf{45}$ $ \mathbf{\bar{40}}$ & $0$ & $-200$ & \\ [-5pt]
10)\phantom{$\Big|$}& $\mathbf{70}$ $\mathbf{15}$ & $0$ & $100$ & \\ [-0pt]
\hline
11)\phantom{$\Big|$}& $\mathbf{5}$ $\mathbf{5}$ & $-9$ & $-4$ & $\lower 0pt\hbox{$\mathbf{ \Sigma \Sigma\bar{\phi}_d}$}$  \\  [-5pt]
12)\phantom{$\Big|$}& $\mathbf{45}$ $\mathbf{5}$ & $75$ & $100$ &  \\  [-5pt]
13)\phantom{$\Big|$}& $\mathbf{70}$ $\mathbf{5}$ & $-225$ & $-200$ &   \\  [-5pt]
14)\phantom{$\Big|$}& $\mathbf{45}_{\uparrow}$ $ \mathbf{45}$ & $285$ & $-60$ & \\ [-5pt]
15)\phantom{$\Big|$}& $\mathbf{45}_{\downarrow}$ $\mathbf{45}$ & $105$ & $20$ & \\ [-5pt]
16)\phantom{$\Big|$}& $\mathbf{5}$ $\mathbf{45}$ & $45$ & $-20$ & \\  [-5pt]
17)\phantom{$\Big|$}& $\mathbf{70}$ $\mathbf{45}$ & $225$ & $-200$ &  \\  [-0pt]
\hline
\end{tabular}
\caption{Operators contributing to $Y^\ell$ and $Y^d$ at ${\cal O}(\epsilon^2)$.}
\end{center}
\label{table:eps2}
\end{table}
\vspace{-0cm}
\begin{table}[htbp]
\begin{center}
\begin{tabular}{|r|c|r|r|c|}
\hline
 \phantom{$\Big|$} &${\cal O}(\epsilon^3)$ & $\frac{Y^\ell}{(\epsilon/\sqrt{60})^3}$ &  $\frac{Y^d}{(\epsilon/\sqrt{60})^3}$ & \\ 
\hline
1)\phantom{$\Big|$}& $\mathbf{10}$ $\mathbf{10}$ $\mathbf{10}$ & $216$ & $1$ & 
$\lower 0pt\hbox{$\mathbf{\bar{\phi}_d \Sigma \Sigma \Sigma}$}$  \\  [-5pt]
2)\phantom{$\Big|$}& $\mathbf{10}$ $\mathbf{10}$ $\mathbf{15}$ & $0$ & $25$ &  \\  [-5pt]
3)\phantom{$\Big|$}& $\mathbf{10}$ $\mathbf{10}$ $\mathbf{\bar{40}}$ & $0$ & $50$ &   \\  [-5pt]
4)\phantom{$\Big|$}& $\mathbf{10}$ $\mathbf{15}$ $ \mathbf{10}$ & $0$ & $25$ & \\ [-5pt]
5)\phantom{$\Big|$}& $\mathbf{10}$ $\mathbf{15}$ $\mathbf{15}$ & $0$ & $-25$ & \\ [-5pt]
6)\phantom{$\Big|$}& $\mathbf{10}$ $\mathbf{\bar{40}}$  $\mathbf{10}$ & $0$ & $50$ & \\  [-5pt]
7)\phantom{$\Big|$}& $\mathbf{10}$ $\mathbf{\bar{40}}_{\uparrow}$ $\mathbf{\bar{40}}$ & $0$ & $-200$ &  \\  [-5pt]
8)\phantom{$\Big|$}& $\mathbf{10}$ $\mathbf{\bar{40}}_{\downarrow}$ $\mathbf{\bar{40}}$ & $0$ & $350$ &  \\  [-5pt]
9)\phantom{$\Big|$}& $\mathbf{15}$ $\mathbf{10}$ $\mathbf{10}$& $0$ & $5$ &  \\ [-5pt]
10)\phantom{$\Big|$}& $\mathbf{15}$  $\mathbf{10}$ $\mathbf{15}$ & $0$ & $125$ &   \\  [-5pt]
11)\phantom{$\Big|$}& $\mathbf{15}$ $\mathbf{10}$ $ \mathbf{\bar{40}}$ & $0$ & $250$ & \\ [-5pt]
12)\phantom{$\Big|$}& $\mathbf{15}$ $\mathbf{15}$ $\mathbf{10}$& $0$ & $-5$ & \\ [-5pt]
13)\phantom{$\Big|$}& $\mathbf{15}$  $\mathbf{15}$ $\mathbf{15}$ & $0$ & $5$ & \\  [-0pt]
\hline
14)\phantom{$\Big|$}& $\mathbf{5}$ $\mathbf{10}$ $\mathbf{10}$ & $-108$ & $2$& 
$\lower 0pt\hbox{$\mathbf{\Sigma \bar{\phi}_d \Sigma \Sigma }$}$ \\ [-5pt]
15)\phantom{$\Big|$}& $\mathbf{5}$ $\mathbf{10}$ $\mathbf{15}$ & $0$ & $50$ &   \\  [-6pt]
16)\phantom{$\Big|$}& $\mathbf{5}$ $\mathbf{10}$ $ \mathbf{\bar{40}}$ & $0$ & $100$ & \\ [-6pt]
17)\phantom{$\Big|$}& $\mathbf{5}$ $\mathbf{15}$ $\mathbf{10}$ & $0$ & $10$ & \\ [-6pt]
18)\phantom{$\Big|$}& $\mathbf{5}$ $\mathbf{15}$ $\mathbf{15}$ & $0$ & $-10$ & \\  [-6pt]
19)\phantom{$\Big|$}& $\mathbf{45}$ $\mathbf{10}$ $\mathbf{10}$ & $540$ & $10$ &  \\  [-6pt]
20)\phantom{$\Big|$}& $\mathbf{45}$ $\mathbf{10}$ $\mathbf{15}$ & $0$ & $250$ &  \\  [-6pt]
21)\phantom{$\Big|$}& $\mathbf{45}$ $\mathbf{10}$ $\mathbf{\bar{40}}$ & $0$ & $500$ &  \\  [-6pt]
22)\phantom{$\Big|$}& $\mathbf{45}$ $\mathbf{\bar{40}}$ $\mathbf{10}$ & $0$ & $-200$ &  \\  [-6pt]
23)\phantom{$\Big|$}& $\mathbf{45}$ $\mathbf{\bar{40}}_{\uparrow}$ $\mathbf{\bar{40}}$ & $0$ & $800$ &  \\  [-6pt]
24)\phantom{$\Big|$}& $\mathbf{45}$ $\mathbf{\bar{40}}_{\downarrow}$ $\mathbf{\bar{40}}$ & $0$ & $-1400$ &  \\  [-6pt]
25)\phantom{$\Big|$}& $\mathbf{70}$ $\mathbf{15}$ $\mathbf{10}$ & $0$ & $-100$ &  \\  [-6pt]
26)\phantom{$\Big|$}& $\mathbf{70}$ $\mathbf{15}$ $\mathbf{15}$ & $0$ & $100$ &  \\  [-0pt]
\hline
\end{tabular}
\caption{Operators contributing to $Y^\ell$ and $Y^d$ at  ${\cal O}(\epsilon^3)$ (continued below).}
\end{center}
\label{table:eps3a}
\end{table}
%
%
\begin{table}[htbp]
\begin{tabular}{|r|c|r|r|c|}
\hline
 \phantom{$\Big|$} &${\cal O}(\epsilon^3)$ & $\frac{Y^\ell}{(\epsilon/\sqrt{60})^3}$ &  $\frac{Y^d}{(\epsilon/\sqrt{60})^3}$ & \\ 
\hline
27)\phantom{$\Big|$}& $\mathbf{5}$ $\mathbf{5}$ $\mathbf{10}$ & $54$ & $4$& 
$\lower 0pt\hbox{$\mathbf{ \Sigma \Sigma \bar{\phi}_d \Sigma}$}$  \\  [-6pt]
28)\phantom{$\Big|$}& $\mathbf{5}$ $\mathbf{5}$ $\mathbf{15}$ & $0$ & $20$ &  \\  [-6pt]
29)\phantom{$\Big|$}& $\mathbf{5}$ $\mathbf{45}$ $\mathbf{10}$ & $-270$ & $20$ &   \\  [-6pt]
30)\phantom{$\Big|$}& $\mathbf{5}$ $\mathbf{45}$ $ \mathbf{\bar{40}}$ & $0$ & $400$ & \\ [-6pt]
31)\phantom{$\Big|$}& $\mathbf{5}$ $\mathbf{70}$ $ \mathbf{15}$ & $0$ & $-200$ & \\ [-6pt]
32)\phantom{$\Big|$}& $\mathbf{45}$ $\mathbf{5}$ $\mathbf{10}$ & $-450$ & $-100$ & \\ [-6pt]
33)\phantom{$\Big|$}& $\mathbf{45}$ $\mathbf{5}$  $\mathbf{15}$ & $0$ & $-500$ & \\  [-6pt]
34)\phantom{$\Big|$}& $\mathbf{45}_{\uparrow}$ $\mathbf{45}$ $\mathbf{10}$ & $-1710$ & $60$ & \\ [-6pt]
35)\phantom{$\Big|$}& $\mathbf{45}_{\downarrow}$ $\mathbf{45}$ $\mathbf{10}$ & $-630$ & $-20$ & \\ [-6pt]
36)\phantom{$\Big|$}& $\mathbf{45}_{\uparrow}$ $\mathbf{45}$ $\mathbf{\bar{40}}$ & $0$ & $-1200$ & \\ [-6pt]
37)\phantom{$\Big|$}& $\mathbf{45}_{\downarrow}$ $\mathbf{45}$ $\mathbf{\bar{40}}$ & $0$ & $400$ & \\ [-6pt]
38)\phantom{$\Big|$}& $\mathbf{45}$ $\mathbf{\bar{50}}$ $\mathbf{\bar{40}}$ & $0$ & $-$ & \\ [-6pt]
39)\phantom{$\Big|$}& $\mathbf{45}$ $\mathbf{70}$ $\mathbf{15}$ & $0$ & $1000$ & \\ [-6pt]
40)\phantom{$\Big|$}& $\mathbf{70}$ $\mathbf{5}$ $\mathbf{10}$ & $1350$ & $200$ &  \\  [-6pt]
41)\phantom{$\Big|$}& $\mathbf{70}$ $\mathbf{5}$ $\mathbf{15}$ & $0$ & $1000$ &  \\  [-6pt]
42)\phantom{$\Big|$}& $\mathbf{70}$ $\mathbf{45}$ $\mathbf{10}$ & $-1350$ & $200$ &  \\  [-6pt]
43)\phantom{$\Big|$}& $\mathbf{70}$ $\mathbf{45}$ $\mathbf{\bar{40}}$ & $0$ & $4000$ &  \\  [-6pt]
44)\phantom{$\Big|$}& $\mathbf{70}_{\uparrow}$ $\mathbf{70}$ $\mathbf{15}$ & $0$ & $-400$ &  \\  [-6pt]
45)\phantom{$\Big|$}& $\mathbf{70}_{\downarrow}$ $\mathbf{70}$ $\mathbf{15}$ & $0$ & $800$ &  \\  [-0pt]
\hline
46)\phantom{$\Big|$}& $\mathbf{5}$ $\mathbf{5}$ $\mathbf{5}$ & $-27$ & $8$ & 
$\lower 0pt\hbox{$\mathbf{ \Sigma \Sigma \Sigma \bar{\phi}_d}$}$  \\  [-5pt]
47)\phantom{$\Big|$}& $\mathbf{45}$ $\mathbf{5}$ $\mathbf{5}$ & $225$ & $-200$ &  \\  [-5pt]
48)\phantom{$\Big|$}& $\mathbf{70}$ $\mathbf{5}$ $\mathbf{5}$ & $-675$ & $400$ &  \\  [-5pt]
49)\phantom{$\Big|$}& $\mathbf{5}$ $\mathbf{45}$ $\mathbf{5}$& $225$ & $-200$ &  \\ [-5pt]
50)\phantom{$\Big|$}& $\mathbf{45}_{\uparrow}$  $\mathbf{45}$ $\mathbf{5}$ & $1425$ & $-600$ &   \\  [-5pt]
51)\phantom{$\Big|$}& $\mathbf{45}_{\downarrow}$ $\mathbf{45}$ $ \mathbf{5}$ & $525$ & $200$ & \\ [-5pt]
52)\phantom{$\Big|$}& $\mathbf{70}$ $\mathbf{45}$ $\mathbf{5}$& $1125$ & $-2000$ & \\ [-5pt]
53)\phantom{$\Big|$}& $\mathbf{5}$  $\mathbf{70}$ $\mathbf{5}$ & $-675$ & $400$ & \\  [-5pt]
54)\phantom{$\Big|$}& $\mathbf{45}$ $\mathbf{70}$ $\mathbf{5}$ & $1125$ & $-2000$ & \\ [-5pt]
55)\phantom{$\Big|$}& $\mathbf{70}_{\uparrow}$ $\mathbf{70}$ $\mathbf{5}$ & $-4725$ & $800$ &   \\  [-5pt]
56)\phantom{$\Big|$}& $\mathbf{70}_{\downarrow}$ $\mathbf{70}$ $ \mathbf{5}$ & $-675$ & $-1600$ & \\ [-5pt]
57)\phantom{$\Big|$}& $\mathbf{5}$ $\mathbf{5}$ $\mathbf{45}$ & $135$ & $40$ & \\ [-5pt]
58)\phantom{$\Big|$}& $\mathbf{45}$ $\mathbf{5}$ $\mathbf{45}$ & $-1125$ & $-1000$ & \\  [-5pt]
59)\phantom{$\Big|$}& $\mathbf{70}$ $\mathbf{5}$ $\mathbf{45}$ & $3375$ & $2000$ &  \\  [-5pt]
60)\phantom{$\Big|$}& $\mathbf{5}$ $\mathbf{45}_{\uparrow}$ $\mathbf{45}$ & $855$ & $120$ &  \\  [-5pt]
61)\phantom{$\Big|$}& $\mathbf{5}$ $\mathbf{45}_{\downarrow}$ $\mathbf{45}$ & $315$ & $-40$ &  \\  [-5pt]
62)\phantom{$\Big|$}& $\mathbf{45}_{\uparrow}$ $\mathbf{45}_{\uparrow}$ $\mathbf{45}$ & $5415$ & $360$ &  \\  [-5pt]
63)\phantom{$\Big|$}& $\mathbf{45}_{\uparrow}$ $\mathbf{45}_{\downarrow}$ $\mathbf{45}$ & $1995$ & $-120$ &  \\  [-5pt]
64)\phantom{$\Big|$}& $\mathbf{45}_{\downarrow}$ $\mathbf{45}_{\uparrow}$ $\mathbf{45}$ & $1995$ & $-120$ &  \\  [-5pt]
65)\phantom{$\Big|$}& $\mathbf{45}_{\downarrow}$ $\mathbf{45}_{\downarrow}$ $\mathbf{45}$ & $735$ & $40$ &  \\  [-5pt]
66)\phantom{$\Big|$}& $\mathbf{70}$ $\mathbf{45}_{\uparrow}$ $\mathbf{45}$ & $4275$ & $1200$ &  \\  [-5pt]
67)\phantom{$\Big|$}& $\mathbf{70}$ $\mathbf{45}_{\downarrow}$ $\mathbf{45}$ & $1575$ & $-400$ &  \\  [-5pt]
68)\phantom{$\Big|$}& $\mathbf{5}$ $\mathbf{70}$ $\mathbf{45}$ & $675$ & $400$ &  \\  [-5pt]
69)\phantom{$\Big|$}& $\mathbf{45}$ $\mathbf{70}$ $\mathbf{45}$ & $-1125$ & $-2000$ &  \\  [-5pt]
70)\phantom{$\Big|$}& $\mathbf{70}_{\uparrow}$ $\mathbf{70}$ $\mathbf{45}$ & $4725$ & $800$ &  \\  [-5pt]
71)\phantom{$\Big|$}& $\mathbf{70}_{\downarrow}$ $\mathbf{70}$ $\mathbf{\bar{45}}$ & $675$ & $-1600$ &  \\  [-0pt]
\hline
\end{tabular}
\caption{Operators contributing to $Y^\ell$ and $Y^d$ at  ${\cal O}(\epsilon^3)$.
\protect\label{table:eps3b}
}
\end{table}

\vfill 

 \newpage 

\null 

 \phantom{Change page for the bibliography} 
 \bigskip
 \phantom{Change page for the bibliography} 

 \vspace{3cm}


\end{document}